%% file: smc.tex
\documentclass[11pt]{article}

\input{header}
\graphicspath{{figures/}}

\newcommand{\blind}{0}

\spacingset{1} \title{{\bf Sequential Monte Carlo for Sampling
    Balanced and Compact Redistricting Plans}\if\blind0\thanks{We
    thank Moon Duchin, Ben Fifield, Greg Herschlag, Mike Higgins,
    Chris Kenny, Jonathan Mattingly, Justin Solomon, and Alex Tarr for
    helpful comments and conversations.  Imai thanks Yunkyu Sohn for
    his contributions at an initial phase of this project.
    Open-source software is available for implementing the proposed
    methodology \citep{redist}.}\fi}

\if\blind0\author{Cory M\MakeLowercase{c}Cartan\thanks{Ph.D. candidate, Department of
    Statistics, Harvard University. 1 Oxford Street, Cambridge 02138. Email:
    \href{mailto:cmccartan@g.harvard.edu}{\texttt{cmccartan@g.harvard.edu}}} 
    \and Kosuke Imai\thanks{Professor, Department of Government and 
    Department of Statistics, Harvard
    University.  1737 Cambridge Street, Institute for Quantitative
    Social Science, Cambridge 02138. Email:
    \href{mailto:imai@harvard.edu}{\texttt{imai@harvard.edu}}, URL:
    \href{https://imai.fas.harvard.edu/}{\tt https://imai.fas.harvard.edu/}}}\fi

\if\blind0\fi
\date{First Draft: July 6, 2020\\
This Draft: \today}

\begin{document}

\maketitle

\pdfbookmark[1]{Title Page}{Title Page}

\thispagestyle{empty}
\setcounter{page}{0}
   
\begin{abstract}
  Random sampling of graph partitions under constraints has become a
  popular tool for evaluating legislative redistricting plans.
  Analysts detect partisan gerrymandering by comparing a proposed
  redistricting plan with an ensemble of sampled alternative plans.
  For successful application, sampling methods must scale to
  maps with a moderate or large number of districts, incorporate realistic legal constraints,
  and accurately and efficiently sample from a selected target
  distribution.  Unfortunately, most existing methods struggle in at
  least one of these areas.  We present a new Sequential Monte Carlo
  (SMC) algorithm that generates a sample of redistricting plans converging
  to a realistic target distribution. Because it draws many plans in 
  parallel, the SMC algorithm can efficiently explore the relevant
  space of redistricting plans better than the existing Markov chain
  Monte Carlo (MCMC) algorithms that generate plans sequentially. Our
  algorithm can simultaneously incorporate several constraints
  commonly imposed in real-world redistricting problems, including
  equal population, compactness, and preservation of administrative
  boundaries.  We validate the accuracy of the proposed algorithm by
  using a small map where all redistricting plans can be enumerated.
  We then apply the SMC algorithm to evaluate the partisan
  implications of several maps submitted by relevant parties in a
  recent high-profile redistricting case in the state of Pennsylvania. 
  We find that the proposed algorithm converges 
  faster and with fewer samples than a comparable MCMC algorithm. 
  Open-source software is available for implementing the proposed methodology.

\end{abstract}

\spacingset{1}
\newpage
\section{Introduction} 

In first-past-the-post electoral systems, legislative districts
serve as the fundamental building block of democratic representation.
In the United States, congressional redistricting, which redraws
district boundaries in each state following the decennial Census, plays a
central role in influencing who is elected and hence what policies are
eventually enacted.  Because the stakes are so high, redistricting has
been subject to intense political battles.  Parties often engage in
\textit{gerrymandering} by manipulating district boundaries in order
to amplify the voting power of some groups while diluting that of
others.

In recent years, the availability of granular data about individual
voters has led to sophisticated partisan gerrymandering attempts that
cannot be easily detected.  At the same time, many scholars have
focused their efforts on developing methods to uncover gerrymandering
by comparing a proposed redistricting plan with a large collection of
alternative plans that satisfy the relevant legal requirements.  A
primary advantage of such an approach over the use of simple summary
statistics is its ability to account for the characteristics of each
state's physical and political geography and state-specific
redistricting rules.

For its successful application, a sampling algorithm for drawing
alternative plans must (1) be efficient enough to scale to maps 
with thousands of geographic units and a moderate or large number of districts, 
(2) simultaneously incorporate a variety of real-world legal 
constraints such as population balance (Section~\ref{subsec:setup}), 
geographical compactness (Section~\ref{subsec:comp}), 
and the preservation of administrative boundaries
(Section~\ref{subsec:adminboundary}), and
(3) ensure these samples are representative of a specific target population, 
against which a redistricting plan of interest can be evaluated.  
Although some have been used in several recent court challenges to 
existing redistricting plans, all existing algorithms run into limitations 
of varying severity with regards to at least one of these three key requirements.

Optimization-based
\citep[e.g.,][]{mehrotra1998,macmillan2001,bozkaya2003,liu2016} and
constructive Monte Carlo
\citep[e.g.,][]{cirincione2000,chen2013,magleby2018} methods can be
made scalable and incorporate many constraints. But they are not
designed to sample from any specific target distribution.  As a
consequence, the resulting plans tend to differ systematically, for
example, from a uniform distribution under certain constraints
\citep{cho2018,fifield2020mcmc,fifield2020enum}. The absence of an
explicit target distribution makes it difficult to interpret the
ensembles generated by these methods and use them for statistical
outlier analysis to detect gerrymandering.

MCMC algorithms \citep[e.g.,][]{mattingly2014,
  wu2015,chikina2017,deford2019,carter2019,fifield2020mcmc,cannon2022spanning} can in
theory sample from a specific target distribution, and incorporate
constraints through the use of an energy function.  In practice,
however, existing algorithms struggle to mix and traverse through a
highly complex sampling space, making scalability difficult and
accuracy hard to prove.  Some of these algorithms make proposals by
flipping precincts at the boundary of existing districts
\citep[e.g.,][]{mattingly2014,fifield2020mcmc}, rendering it difficult
or even impossible to transition between points in the state space, especially
as more constraints are imposed.  
More recent algorithms by \citet{deford2019} and \citet{carter2019} use
spanning trees to make their proposals, and this has allowed these algorithms 
to yield greater moves and substantially improve mixing. 
Yet recent theoretical results suggest that even these larger moves may not be 
enough to traverse the entire state space, and therefore may fail to converge to the correct 
distribution, if a realistic population balance constraint is imposed
\citep{akitaya2022reconfiguration}.

We contribute to the ongoing scholarly efforts
to address the above three key challenges by developing a new
Sequential Monte Carlo (SMC) algorithm, based on a similar but not
identical spanning tree construction to \citet{deford2019} and 
\citet{carter2019} (see Sections~\ref{sec:goal}~and~\ref{sec:algo}).
Like MCMC algorithms, the SMC algorithm generates samples which approximate
the target distribution arbitrarily well as the sample size increases.
But like constructive Monte Carlo methods, the SMC algorithm draws many 
separate plans from scratch, rather than tweaking a single plan sequentially.
This approach is better suited to the large discrete state space with a 
multimodal target distribution that characterizes redistricting problems.
For example, in cases where existing MCMC proposals render the state space disconnected, the
SMC algorithm can still converge to the target distribution.
As we demonstrate in
Sections~\ref{sec:validation}~and~\ref{sec:pa-study} (see also Appendix~\ref{app:fl50}), this
sampling approach translates to faster convergence and smaller standard errors
for a given computational budget.
For larger and more complex redistricting sampling problems, the SMC algorithm can
be easily parallelized to facilitate efficient computation. 

The proposed algorithm proceeds by splitting off one district at a
time, building up the redistricting plan piece by piece (see
Figure~\ref{fig:seq-overview} for an illustration).  Each split is
accomplished by drawing a spanning tree and removing one edge, which
splits the spanning tree in two. We also extend the SMC algorithm so
that it preserves administrative boundaries and certain geographical
areas as much as possible, which is another common constraint
considered in many real-world redistricting cases. \if\blind0 An
open-source software package, \texttt{redist}, is available for implementing 
the proposed algorithm \citep{redist}.\fi

The SMC algorithm is not without limitations.
Like existing MCMC approaches, the SMC algorithm only guarantees convergence
to the target distribution as the sample size approaches infinity. 
Additionally, because the SMC algorithm involves repeated resampling, with
a finite number of samples it can suffer from particle system collapse
\citep{liu2001}, 
significantly increasing sampling variability.
Thus, it is important to understand the limitations of the proposed algorithm
in finite samples, especially when dealing with large maps and many districts. 
In Section~\ref{subsubsec:diag}, we provide a set of diagnostics which
can be used in practice to help identify if more samples are needed to reach 
convergence or if the constraints imposed by an analyst are too
strong.

In Section~\ref{sec:validation}, we validate the SMC algorithm using a
small map for which all potential redistricting plans can be
enumerated \citep{fifield2020enum}. 
We demonstrate that the proposed algorithm samples accurately from a 
realistic target distribution, and that our proposed diagnostics are a good 
proxy for total sampling error, which is generally unobservable.
Section~\ref{sec:pa-study} applies the SMC algorithm to
the 2011 Pennsylvania congressional redistricting, and also compares 
its performance on this problem with an MCMC algorithm with the same
transition kernel as the proposed SMC algorithm.
We find that the SMC algorithm samples more efficiently than the MCMC approach.  
Section~\ref{sec:conclusion} concludes and discusses directions for future work.

\section{The 2011 Pennsylvania Congressional Redistricting}
\label{sec:pa-overview}

We study the 2011 Pennsylvania congressional redistricting because it
illustrates the salient features of the redistricting problem.  We
begin by briefly summarizing the background of this case and then
explain the role of sampling algorithms used in the expert witness reports.

\subsection{Background}
Pennsylvania lost a seat in Congress during the reapportionment of the
435 U.S. House seats following the 2010 Census.  In Pennsylvania, the
General Assembly, which is the state's legislative body, draws new
congressional districts, subject to gubernatorial veto.  At the time,
the General Assembly was controlled by Republicans, and Tom Corbett,
also a Republican, served as governor.  In the 2012 election, which
took place under the newly adopted 2011 districting map, Democrats won
5 seats while Republicans took the remaining 13.  Under the previous
plan, the split was 7--12.

In June 2017, the League of Women Voters of Pennsylvania
filed a lawsuit alleging that the 2011 plan adopted by the Republican
legislature violated the state constitution by diluting the political
power of Democratic voters. The case worked its way through the state
court system, and on January 22, 2018, the Pennsylvania Supreme Court
issued its ruling, writing that the 2011 plan ``clearly, plainly and
palpably violates the Constitution of the Commonwealth of
Pennsylvania, and, on that sole basis, we hereby strike it as
unconstitutional.''
\citep{2018league}.

The court ordered that the General Assembly adopt a remedial plan and
submit it to the governor, who would in turn submit it to the court,
by February 15, 2018. In its ruling, the court laid out specific requirements
that had to be satisfied by all proposed plans:
\begin{quote}
  composed of compact and contiguous territory; 
  as nearly equal in population as practicable; and which do not divide any
  county, city, incorporated town, borough, township, or ward, except
  where necessary to ensure equality of population.
\end{quote}

\begin{figure}[t]
  \spacingset{1}
    \begin{subfigure}{.5\textwidth}
      \centering
      \includegraphics[width=2.6in]{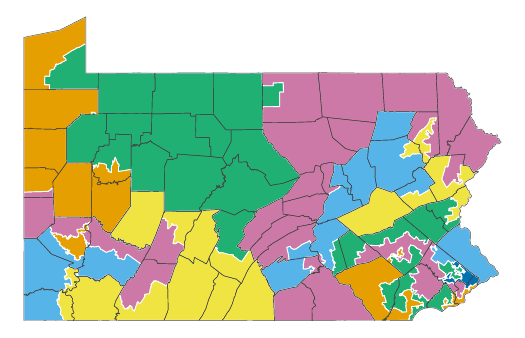}
      \subcaption{2011 General Assembly map}
    \end{subfigure}
    \begin{subfigure}{.5\textwidth}
      \centering
      \includegraphics[width=2.6in]{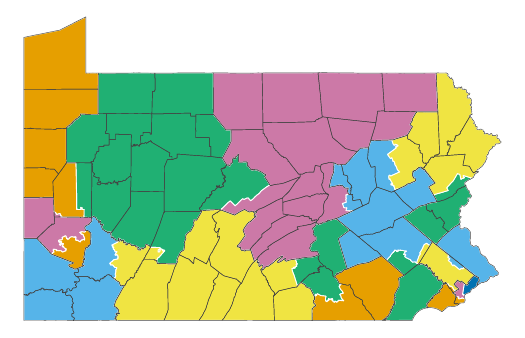}
      \subcaption{2018 Pennsylvania Supreme Court map}
    \end{subfigure}
    \caption{Comparison of the 2011 map drawn by the General Assembly
      and the final map imposed by the Supreme court in 2018. County lines are
      shown in dark gray, and district boundaries that do not coincide with
      county boundaries are in white.}
    \label{fig:pa-maps}
\end{figure}

The leaders of the Republican Party in the General Assembly drew a new
map, but the Democratic governor, Tom Wolf, refused to submit it to
the court, claiming that it, too, was an unconstitutional gerrymander.
Instead, the court received remedial plans from seven parties: the
petitioners, the League of Women Voters; the respondents, the
Republican leaders of the General Assembly; the governor, a Democrat;
the lieutenant governor, also a Democrat; the Democratic Pennsylvania
House minority leadership; the Democratic Pennsylvania Senate minority
leadership; and the intervenors, which included Republican party
candidates and officials.  Ultimately, the Supreme Court drew its own
plan and adopted it on February 19, 2018, arguing that it was
``superior or comparable to all plans submitted by the parties."
Figure~\ref{fig:pa-maps} shows the remedial plan created by the
Supreme Court as well as the 2011 map adopted by the General Assembly,
which were found on the court's case page.

The constraints explicitly laid out by the court, as well as the
numerous remedial plans submitted by the parties, make the 2011
Pennsylvania redistricting a useful case study that evaluates
redistricting plans.

\subsection{The Role of Sampling Algorithms}

The original finding that the 2011 General Assembly plan was a
partisan gerrymander was in part based on different outlier analyses
performed by two academic researchers, Jowei Chen and Wesley Pegden,
who served as the petitioner's expert witnesses.
Chen randomly generated two sets of 500 redistricting plans according
to a constructive Monte Carlo algorithm based on \citet{chen2013}.  He
considered population balance, contiguity, compactness, avoiding
county and municipal splits, and, in the second set of 500, avoiding
plans that placed more than two incumbents in the same district (at least one
pair of incumbents in the same district was necessary, given that Pennsylvania 
lost a seat from 2000 to 2010). 
Pegden ran a reversible Markov chain similar to that used in the MCMC 
algorithm of \citet{mattingly2014} for one trillion steps, and computed 
upper bounds of $p$-values using the method of \citet{chikina2017}. 
This method was also used in a follow-up analysis by Moon Duchin, who
served as an expert for Governor Wolf \citep{duchin2018}. Both
petitioner experts concluded that the 2011 plan was an extreme outlier
according to compactness, county and municipal splits, and the number
of Republican and Democratic seats implied by statewide election
results.

The respondents also retained an expert academic witness, Wendy Tam Cho, who 
directly addressed the sampling-based analyses of Chen and Pegden. 
Cho criticized Chen's analysis for not sampling from a specified
target distribution.  She also criticized Pedgen's analysis by arguing
that his Markov chain only made local explorations of the space of
redistricting plans, and could not therefore have generated a
representative sample of all valid plans, though the $p$-values computed
using the \citet{chikina2017} method explicitly do not require mixing of the 
Markov chain \citep[see also][and \citet{chikina2019reply}]{cho2019}.
We do not directly examine the intellectual merits of the specific arguments
put forth by the expert witnesses.
However, these methodological debates are also relevant
for other cases where simulation algorithms have been extensively used
by expert witnesses (e.g., Rucho v. Common Cause (2019); Common Cause
v. Lewis (2019); Covington v. North Carolina (2017); Harper v. Lewis
(2020)), and highlight the difficulties in practically applying 
existing sampling algorithms to actual redistricting problems.

First, the distributions that some of these algorithms sample from are not 
made explicit, leaving open the possibility that the generated ensemble is 
systematically different from the true set of all valid plans.
Second, even when the distribution is known, MCMC algorithms used to 
sample from it may be prohibitively slow to mix and cannot yield a 
representative sample.
These challenges motivate us to design an algorithm that accurately
samples from a specific target distribution and incorporates most common 
redistricting constraints, while being efficient and scalable.

\section{Sampling Balanced and Compact Districts} \label{sec:goal}


\subsection{The Setup} \label{subsec:setup}

Redistricting plans are ultimately aggregations of geographic units
such as counties, voting precincts, or Census blocks. The usual
requirement that the districts in a plan be contiguous necessitates
consideration of the spatial relationship between these units. The
natural mathematical structure for this consideration is a graph
$G=(V,E)$, where $V=\{v_1, v_2, \ldots, v_m\}$ consists of $m$ nodes
representing the geographic units of redistricting and $E$ contains
edges connecting units which are legally adjacent.

A labeled redistricting plan on $G$ consists of $n$ districts, where each 
district is a collection of nodes. 
A labeled plan is described by a function $\xi:V\to\{1,2,\dots,n\}$, 
where $\xi(v)=i$ implies that node $v$ is in district $i$. 
In practice, we will be interested in 
\textit{unlabeled} plans, since the assignment of sets of precincts to labels is
arbitrary and has no impact on the real-world aspects of the district such as its
population, demographic composition, or partisan lean. Define an equivalence relation 
by $\xi_1\cong\xi_2$ if there exists a permutation $\sigma$ such that 
$\xi_1(v)=\sigma(\xi_2(v))$ for all $v$. Then, an unlabeled redistricting 
plan can be viewed as an equivalence class under $\cong$, denoted $[\xi]$;
nothing  in what follows will depend on the particular labeled
plan representative $\xi$.

We let $V_i(\xi)$ and $E_i(\xi)$ denote
the nodes and edges contained in district $i$ under a given
redistricting plan $\xi$, so $G_i(\xi)=(V_i(\xi),E_i(\xi))$ represents
the induced subgraph that corresponds to district $i$ under the plan.
We suppress the dependence on $\xi$ when it is clear from context,
writing $G_i=(V_i,E_i)$.  Since each node belongs to only one
district, we have $V = \bigcup_{i=1}^n V_i(\xi)$ and
$V_i(\xi) \bigcap V_{i^\prime}(\xi) = \empty$ for any redistricting
plan $\xi$.  In addition, we require that each district be contiguous,
i.e., that $G_i$ is a connected graph, for all $i$.

Beyond connectedness, redistricting plans are always
required to have roughly equal population in every district.  To
formalize this requirement, let $\pop(v)$ denote the population of
node $v$.  Then, the population of a district $i$ may be written as
$\pop(V_i(\xi)) \ \dfeq \ \sum_{v\in V_i(\xi)}\pop(v)$.
We quantify the discrepancy between a given plan and the ideal of equal 
population in every district by the \textit{maximum population deviation}, 
\begin{equation*}
    \dev(\xi) \ \dfeq \ \max_{1 \le i\le n}
    \abs{\frac{\pop(V_i)}{\pop(V)/n} - 1},
\end{equation*}
where $\pop(V)$ is the total population. 
Some courts and states have imposed hard maximums on this quantity, e.g.,
$\dev(\xi)\le D=0.05$ for state legislative redistricting \citep{ncslcriteria}.

The proposed algorithm samples plans by way of spanning trees on each
district, i.e., subgraphs of $G_i(\xi)$ which contain all vertices, no
cycles, and are connected.  Let $T_i$ represent a spanning tree for
district $i$ whose vertices and edges are given by $V_i(\xi)$ and a
subset of $E_i(\xi)$, respectively.  The collection of spanning trees
from all districts together form a spanning forest.
Each node belongs to one spanning tree in the forest, and this assignment 
corresponds to a redistricting plan. However, a single redistricting plan may 
correspond to multiple spanning forests because each district may admit more 
than one spanning tree.

For a given redistricting plan, we can compute the exact number of
spanning forests in polynomial time using the determinant of a
submatrix of the graph Laplacian, according to the Matrix Tree Theorem
of Kirchhoff (see \cite{tutte1984}).  Thus, for a graph $H$, if we let
$\tau(H)$ denote the number of spanning trees on the graph, we can
represent the number of spanning forests that correspond to a
redistricting plan $\xi$ as 
$\tau(\xi) \dfeq \prod_{i=1}^n \tau(G_i(\xi))$.
This fact will play an important role in the definition of our
sampling algorithm and its target distribution.

\subsection{The Target Distribution}

The algorithm is designed to sample an unlabeled plan $[\xi]$ with probability 
\begin{equation} \label{eq:tgt}
    \pi([\xi]) \propto \exp\{-J(\xi)\}\tau(\xi)^{\rho}
    \ind_{\{\xi\text{ connected}\}}\ind_{\{\dev(\xi)\le D\}},
\end{equation}
where the indicator functions ensure that the plans meet population
balance and connectedness criteria, $\tau(\xi)$ measures the
compactness of the districts in $\xi$ (see Section~\ref{subsec:comp}),
and $J$ encodes additional constraints on the types of plans
preferred.  As done in Section~\ref{sec:pa-study}, we often use a
reasonably strict population constraint such as $D=0.001$ and $D=0.005$. The
parameter $\rho\in \R_0^+$ is chosen to control the compactness of the
generated plans.

This target distribution $\pi$ has both substantive and theoretical
justifications.  First, it incorporates two constraints which are almost
always present in real-world redistricting: the support of $\pi$ is restricted to 
contiguous plans and those that meet a population deviation threshold. 
Second, it represents the unique maximum entropy distribution on the set of
redistricting plans satisfying these two universal constraints and
the moment conditions implied by the other constraints, i.e.,
$\E_\pi[\log\tau(\xi)] = \mu_\tau$ and $\E_\pi[J(\xi)] = \mu_J$ for
some constants $\mu_\tau$ and $\mu_J$ \citep[see][Theorem. 12.1.1,
originally of Boltzmann]{cover2006}.

Thus, our target distribution ensures that all plans meet contiguity
and population requirements, and \textit{on average} satisfy a
compactness standard as well as any other additional constraints
(through the function $J$).  It is no surprise, therefore, that this
class of target distributions has been used by other work developing
redistricting sampling algorithms
\citep{herschlag2017,fifield2020mcmc}.

The generality of the additional constraint function $J$ is
intentional, as its exact form and number imposed on the redistricting
process varies by state and by the type of districts being drawn; any
type of constraint may be incorporated by choosing a $J$ which is
small for preferred plans and large otherwise.  For example, a
preference for plans close to an existing plan $\xi_{\text{sq}}$ may
be encoded as
\begin{align*}
    J_{\text{sq}}(\xi) & = -\frac{\beta}{\log n}\mathrm{VI}(\xi,\xi_{\text{sq}}) 
    \dfeq -\frac{\beta}{2\log n} \sum_{i,j=1}^n
    \tfrac{P_{ij}}{\pop(V)}
    \qty(\log(\tfrac{P_{ij}}{\pop(V_j(\xi_{\text{sq}}))})
    + \log(\tfrac{P_{ij}}{\pop(V_j(\xi))})),
    \numberthis \label{eqn:varinfo}
\end{align*}
where $\beta\in\R^+$ controls the strength of the constraint and 
$P_{ij}=\pop(V_i(\xi)\cap V_j(\xi_{\text{sq}}))$ is the population shared 
between district $i$ of $\xi$ and $j$ of $\xi_{\text{sq}}$. The
function ${\rm VI}(\cdot,\cdot)$ represents the variation of
information (also known as the shared information distance), which is
the difference between the joint entropy and the mutual information of
the distribution of population over the new districts $\xi$ relative
to the existing districts $\xi_{sq}$ \citep{cover2006}. When $\xi$ is
any relabeling of $\xi_{\text{sq}}$, then $J_{\text{sq}}(\xi)=0$.  In
contrast, when $\xi$ evenly splits the nodes of each district of
$\xi_{\text{sq}}$ among the districts of $\xi$, then
$J_{\text{sq}}(\xi)=\beta$. This distance measure will prove useful
later in quantifying the diversity of a sample of redistricting plans
\citep[see also][]{guth2020three}.

There exist other formulations of constraints, and considerations in
choosing a set of weights that balance constraints against each other
\citep[see e.g.,][]{bangia2017,herschlag2017,fifield2020mcmc}.  Here,
we focus on sampling from the broad class of distributions
characterized by Equation~\eqref{eq:tgt}, which have been used in
other work; we do not address the important but separate problem of
picking a specific instance of this class for a given redistricting
problem.

The flexibility of $J$ can be deceptive, however.
The algorithm operates efficiently only when the additional
constraints imposed by $J$ are not too severe.  Even a small number of
strong constraints incorporated into $J$ can dramatically limit the
number of valid plans and considerably complicate the process of
sampling \citep{chatterjee2018sample}.
The Markov chain algorithms developed to date partially
avoid this problem by moving toward maps with lower $J$ over a number
of steps, but in general including more constraints makes it even more
difficult to transition between valid redistricting plans.  Approaches
such as simulated annealing \citep{bangia2017,herschlag2017} and
parallel tempering \citep{fifield2020mcmc} have been proposed to
handle multiple constraints, but these can be difficult to calibrate
in practice and provide few, if any, theoretical guarantees.

In practice, we usually find that the most stringent constraints are
those involving population deviation, compactness, and administrative
boundary splits.  As shown later, we address this issue by designing our
algorithm to directly satisfy these constraints.  Weak additional
constraints do not generally have a substantial effect on the sampling
efficiency, though there are exceptions.  Monitoring the distribution of
the weights and the overall sampling efficiency is crucial to obtaining
a good sample, as we discuss later.

\subsection{Spanning Forests and Compactness} \label{subsec:comp}

One common redistricting requirement is that districts be
geographically compact, though nearly every state leaves this term
undefined.  Dozens of numerical compactness measures have been
proposed, with the Polsby--Popper score \citep{polsby1991} perhaps the
most popular. Defined as the ratio of a district's area to that of a
circle with the same perimeter as the district, the Polsby--Popper
score is constrained to $[0,1]$, with higher scores indicating more
compactness.

Other scholars have proposed a graph-theoretic
measure known as \textit{edge-cut compactness} \citep{dube2016,
  deford2019}.  This measure counts the number of edges that must be
removed from the original graph to partition it according to a given
plan. Formally, it is defined as
\begin{equation*}
  \rem(\xi) \ \dfeq \ 1 - \frac{\sum_{i=1}^n |E_i(\xi)|}{|E(G)|},
\end{equation*}
where we have normalized to the total number of edges.

Plans that involve cutting many edges will necessarily have long
internal boundaries, driving up their average district perimeter (and
driving down their Polsby--Popper scores), while plans that cut as few
edges as possible will have relatively short internal boundaries and
much more compact districts.  Additionally, given the high density of
voting units in urban areas, plans which cut fewer edges will tend to
avoid drawing district lines through the heart of these urban areas.
This has the welcome side effect of avoiding splitting cities and
towns, and in doing so helping to preserve ``communities of interest,"
another common redistricting consideration. 

Empirically, this graph-based compactness measure is highly correlated
with $\log\tau(G) - \log\tau(\xi)$ (we often observe a
correlation in excess of 0.99).  It is difficult to precisely
characterize this relationship except in special cases because
$\tau(\xi)$ is calculated as a matrix determinant \citep{mckay1981}.
However, this quantity is strongly controlled by the product of the
degrees of each node in the graph, $\prod_{i=1}^m \deg(v_i)$
\citep{kostochka1995}.  Removing an edge from a graph decreases the
degree of the vertices at either end by one, so we would expect
$\log\tau(G)$ to change by approximately
$2\{\log \bar d-\log(\bar d-1)\}$ with this edge removal, where
$\bar d$ is the average degree of the graph. This implies a linear
relation
$\log\tau(G)-\log\tau(\xi) \approx \rem(\xi)\cdot 2\{\log \bar
d-\log(\bar d-1)\}$, and hence 
$\tau(\xi)^\rho \approx C_1\exp(-C_2\,\rho\rem(\xi))$,
where $C_1$ and $C_2$ are some constants depending on the details of the map. 

As a result, a greater value of $\rho$ in the target distribution
corresponds to a preference for fewer edge cuts and therefore a
redistricting plan with more compact districts.  This and the
considerations given in the literature \citep{dube2016,deford2019}
suggest that the target distribution in Equation~\eqref{eq:tgt} with
$\rho=1$ (or another positive value) is a good choice for sampling
compact districts. The choice of $\rho=1$ is computationally convenient,
as it allows us to avoid calculating $\tau(\xi)$ as part of sampling (an
asymptotic bottleneck), and yet usually produces satisfactorily compact
districts.  Of course, if another compactness metric is
desired, one can simply set $\rho=0$ and incorporate the alternative
metric into $J$. This will preserve the algorithm's efficiency to the
extent that the alternative metric correlates with the edge-removal
measure of compactness.
Setting $\rho=0$ by itself, however, will make sampling 
intractable in most cases, just as choosing an extreme $J$ will.

\section{The Proposed Algorithm} \label{sec:algo}

The proposed algorithm samples redistricting plans by sequentially
drawing districts over $n-1$ iterations of a splitting procedure.
This is fundamentally different from existing MCMC approaches, which
change an existing plan according to some transition kernel. The
iterations of the proposed algorithm are from district to district
within a single plan whereas the iterations in an MCMC algorithm are
from plan to plan.

Our algorithm begins by partitioning the original graph
$G=(V, E)=(\widetilde V_0, \widetilde E_0)=\widetilde{G}_0$ into two
induced subgraphs: $G_1=(V_1, E_1)$, which will constitute a district
in the final map, and the remainder of the graph
$\widetilde{G}_1=(\widetilde{V}_1, \widetilde{E}_1)$, where
$\widetilde{V}_1=V\setminus V_1$ and $\widetilde{E}_1$ consists of all
the edges between vertices in $\widetilde{V}_1$.  Next, the algorithm
takes $\widetilde{G}_1$ as an input graph and partitions it into two
induced subgraphs, one which will become a district $G_2$ and the
remaining graph $\widetilde{G}_2$.  The algorithm repeats the same
splitting procedure until the final $(n-1)$-th iteration whose two
resulting partitions, $G_{n-1}$ and $\widetilde{G}_{n-1} =G_n$, become
the final two districts of the redistricting plan.

\begin{figure}[t!]
  \centering \spacingset{1}
  \begin{subfigure}[b]{0.23\textwidth}
    \centering
    \includegraphics[width=1.5in]{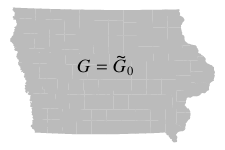}	
    \caption{Initial map}
  \end{subfigure}
  \begin{subfigure}[b]{0.23\textwidth}
    \centering
    \includegraphics[width=1.5in]{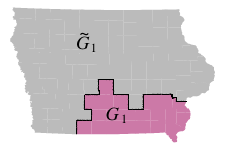}	
    \caption{Iteration 1}
  \end{subfigure}
  \begin{subfigure}[b]{0.23\textwidth}
    \centering
    \includegraphics[width=1.5in]{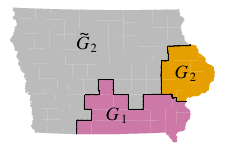}	
    \caption{Iteration 2}
  \end{subfigure}
  \begin{subfigure}[b]{0.23\textwidth}
    \centering
    \includegraphics[width=1.5in]{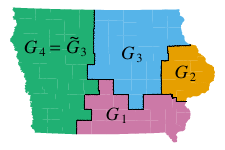}	
    \caption{Iteration 3}
  \end{subfigure}
  \caption{The sequential splitting procedure applied to the state of
    Iowa, where four congressional districts are created at the county
    level.}
  \label{fig:seq-overview}
\end{figure}

Figure~\ref{fig:seq-overview} provides an illustration of this
sequential procedure. To sample a large number of
redistricting plans from the target distribution given in
Equation~\eqref{eq:tgt}, at each iteration, the algorithm samples many
candidate partitions, discards those which fail to meet the population
constraint, and then resamples a certain number of the remainder
according to importance weights, using the resampled partitions at the
next iteration.  The rest of this section explains the details of the
proposed algorithm.

\subsection{The Splitting Procedure}
\label{sub:split}

We first describe the splitting procedure, which is similar to the
merge-split Markov chain proposals of \cite{deford2019} and
\cite{carter2019}.  It proceeds by drawing a random spanning tree $T$,
identifying the $k_i$ most promising edges to cut within the tree, and
selecting one such edge at random to create two induced subgraphs.
Spanning trees are an attractive way to split districts, as the removal
of a single edge induces a partition with two connected components, and
spanning trees can be sampled uniformly \citep{wilson1996}.

\begin{algorithm}[ht]  \spacingset{1}
    \caption{Splitting procedure to generate one district}
    \label{algo:split}
    \textit{Input}: initial graph $\widetilde{G}_{i-1}$ 
    and a parameter $k_i\in\Z^+$. 
    \begin{enumerate}[(a)]
    \item Draw a single spanning tree $T$ on $\widetilde{G}_{i-1}$
        uniformly from the set of all such trees using Wilson's algorithm.
    \item Each edge $e\in E(T)$ divides $T$ into two components,
      $T_e^{(1)}$ and $T_e^{(2)}$. For each edge, compute the following
      population deviation for the two districts that would be induced by
      cutting $T$ at $e$,
      \begin{equation*}
        d_e^{(1)} \ = \  \abs{\frac{\sum_{v \in T_e^{(1)}}\pop(v)}{\pop(V)/n}-1}
        \qand
        d_e^{(2)} \ = \ \abs{\frac{\sum_{v \in T_e^{(2)}}\pop(v)}{\pop(V)/n}-1}.
      \end{equation*}
      Let $d_e=\min \{d_e^{(1)}, d_e^{(2)}\}$, and index the edges in
      ascending order by this quantity, so that we have
      $d_{e_1}\le d_{e_2}\le \cdots\le d_{e_{m_i-1}}$, where $m_i=|\tilde V_{i-1}|$. 
    \item Select one edge $e^*$ uniformly from the top $k_i$ edges,
      $\{e_1,e_2,\dots, e_{k_i}\}$, and remove it from $T$, creating a
      spanning forest $(T_{e^*}^{(1)}, T_{e^*}^{(2)})$ which induces a
      partition $(G_i^{(1)}, G_i^{(2)})$.
    \item If $d_{e^*}^{(1)}\le d_{e^*}^{(2)}$, i.e., if $T_{e^*}^{(1)}$
      induces a district that is closer to the optimal population than
      $T_{e^*}^{(2)}$ does, set $G_i=G_i^{(1)}$ and $\widetilde{G}_i=G_i^{(2)}$;
      otherwise, set $G_i=G_i^{(2)}$ and $\widetilde{G}_i=G_i^{(1)}$.
    \end{enumerate}
\end{algorithm}

As part of the full sampling procedure (Algorithm~\ref{algo:smc}), after
splitting, we check that the population of the new district
$G_i$ falls within the bounds $[P_i^-, P_i^+]$ where
\begin{align*}
    P_i^- &= \max\qty{\frac{\pop(V)}{n}(1-D),\ 
        \pop(\widetilde{V}_{i-1})-\frac{n-i}{n}\pop(V)(1+D)} \qand \\
    P_i^+ &= \min\qty{\frac{\pop(V)}{n}(1+D),\ 
        \pop(\widetilde{V}_{i-1})-\frac{n-i}{n}\pop(V)(1-D)}.
\end{align*}
These bounds also ensure that it will be possible for future iterations to
generate valid districts out of $\widetilde{G}_i$. 
If $\pop(V_i)\not\in[P_i^-,P_i^+]$, then the entire redistricting plan is
rejected and the sampling process begins again.
While the rate of rejection varies by map and by iteration, we generally encounter
acceptance rates at each iteration between 5\% and 30\%, which are not
so low as to make sampling from large maps intractable.
Algorithm~\ref{algo:split} details the steps of the splitting
procedure, where at the first iteration we take $\widetilde{G}_0 = G$.

\subsection{The Sampling Probability}

The above sequential splitting procedure does not generate plans from
the target distribution $\pi$. We denote the sampling
measure by $q$, and write the sampling probability for a given
connected plan $\xi$ at iteration $i$ as $q(G_i\mid\widetilde{G}_{i-1})$,
since each new district $G_i$ depends only
on the leftover map area $\widetilde{G}_{i-1}$ from the previous
iteration. This probability can be written as the probability that we
cut an edge along the boundary of the new district, integrated over all
spanning trees which could be cut to form the district, i.e.,
\begin{equation}\label{eq:rhs-split-pr}
    q(G_i\mid\widetilde{G}_{i-1})
    = \sum_{T\in \mathcal{T}(\widetilde{G}_{i-1})} q(G_i\mid T)\, 
    \tau(\widetilde{G}_{i-1})^{-1},
\end{equation}
where $\mathcal{T}(\cdot)$ represents the set of all spanning trees of a
given graph, and we have relied on the fact that Wilson's algorithm
draws spanning trees uniformly. 

The key 
is that for certain choices of $k_i$ (the number of edges considered
to be cut at iteration $i$), the probability that an edge is cut 
is independent of the trees that are drawn.  Let $ok(T)$ represent the
number of edges on any spanning tree $T$ that induce balanced partitions
with population deviation below $D$, i.e.,
\begin{equation*}
  ok(T) \ \dfeq \ |\{e\in E(T): d_e\le D\}|.
\end{equation*}
Then define
$K_i\dfeq \max_{T\in \mathcal{T}(\widetilde{G}_{i-1})} ok(T)$, the
maximum number of such edges across all spanning trees.  Furthermore,
let $\C(G, H)$ represent the set of edges joining nodes in a subgraph $G$
to nodes in a subgraph $H$. We have the following result for the splitting
probability for new districts whose populations lie inside the bounds
defined above (Appendix~\ref{app:proofs} for the proof).
\begin{restatable}{lemma}{splitprob}\label{lem:splitprob}\spacingset{1}
    The probability of splitting a \emph{valid} new district $G_i$
    from an existing area $\widetilde{G}_{i-1}$ using
    Algorithm~\ref{algo:split} with parameter $k_i\ge K_i$ is
    \begin{align*}
        q(G_i\mid\widetilde{G}_{i-1},\pop(V_i)\in[P_i^-,P_i^+])
        &\ = \ \frac{\tau(G_i)\tau(\widetilde{G}_i)}{\tau(\widetilde{G}_{i-1})k_i} 
            |\C(G_i, \widetilde{G}_i)|.
            \numberthis \label{eq:stage}
    \end{align*}
\end{restatable}

\subsection{Sequential Monte Carlo}
\label{subsec:SMC}

We follow a sequential Monte Carlo approach \citep{doucet2001,liu2001}
to generate draws from the target distribution, rather than simply
performing $n-1$ iterations of Algorithm~\ref{algo:split} and
resampling or reweighting at the final stage. A sequential approach is
also useful in operationalizing the rejection procedure to enforce the
population constraint.

The proposed procedure is presented as Algorithm~\ref{algo:smc}.  The
algorithm is governed by a parameter $\alpha\in(0,1]$, which has no
effect on the target distribution nor the asymptotic accuracy of the
algorithm. Rather, $\alpha$ may be adjusted to maximize the efficiency of
sampling. To generate $S$ redistricting plans, at each iteration of
the splitting procedure $i\in\{1,2,\dots,n-1\}$, we resample and split
the existing plans one at a time, rejecting those which do not meet
the population constraints, until we obtain $S$ new plans for the next
iteration.  This rejection process can be viewed as a form of partial
rejection control \citep{liu1998rejection,liu2001}, or a version of
the AliveSMC algorithm \citep{legland2005sequential, peters2012sequential}.

The weights at each stage serve three purposes. The first is to account for the
extraneous $|\C(G_i,\widetilde{G}_i)|$ term which appears in the
splitting probability $q(G_i\mid\widetilde{G}_{i-1},\pop(V_i)\in[P_i^-,P_i^+])$ 
but not in the target distribution $\pi$. The second is to account for differences 
in the compactness parameter $\rho$; the splitting procedure generates plans 
according to $\rho=1$, but this may be different from the target value.
The third purpose is to adjust for the imbalances between the labeled 
plans generated by the splitting procedure and 
the unlabeled plans which are the target of sampling. While there are $n!$ 
labeled plans corresponding to every unlabeled plan, not every labeled plan may
be sampled according to the sequential procedure. In fact, the number of labeled 
plans  which are in the support of the SMC proposal distribution varies from 
one unlabeled plan to another. It is this imbalance which necessitates 
an additional correction. 

For a labeled plan $\xi$, let $G/\xi$ denote the district-level quotient 
graph corresponding to the plan; i.e., the nodes are districts $\{1,\dots,n\}$
and edges connect adjacent districts. Notice that the splitting procedure
ensures that the leftover area after each split, $\widetilde{G}_i$, is a 
connected graph. This means that for every $1\le i\le n-1$, the subgraph 
of $G/\xi$ induced by the vertices $\{i+1,i+2,\dots,n\}$ (the districts 
drawn after split $i$) is connected. We call any labeled plan $\xi$ a
\textit{sequentially valid labeling} if this property holds, and denote by 
$\psi([\xi])$ the number of sequentially valid labelings corresponding to an 
unlabeled plan $[\xi]$. The SMC weights incorporate $\psi([\xi])$ 
as a correction term. Calculation of $\psi([\xi])$ is discussed below in 
Section~\ref{subsubsec:psi}.

\begin{algorithm}[t!]
    \caption{Sequential Monte Carlo (SMC) Algorithm}
    \label{algo:smc}
    \spacingset{1}
    \textit{Input}: graph $G$ to be split into $n$ districts, target
    distribution parameters $\rho\in\R^+_0$ and constraint function
    $J$, and sampling parameters $\alpha\in (0,1]$ and $k_i\in\Z^+$,
    with $i\in\{1,2,\dots,n-1\}$.
    \begin{enumerate}[(a)]
    \item Generate an initial set of $S$ plans
        $\{\widetilde{G}_0^{(1)},\widetilde{G}_0^{(2)},\dots,\widetilde{G}_0^{(S)}\}$
        and corresponding weights $\{w_0^{(1)},w_0^{(2)},\dots,w_0^{(S)}\}$,
        where each $\widetilde{G}_0^{(j)}\dfeq G$ and $w_0^{(j)}=1$.
    \item For each splitting iteration $i\in\{1,2,\dots,n-1\}$:
        \begin{enumerate}[(1)]
        \item Until there are $S$ valid plans: \begin{enumerate}[(i)]
                \item Sample a partial plan $\widetilde{G}_{i-1}$ from
                    $\{\widetilde{G}_{i-1}^{(1)},\widetilde{G}_{i-1}^{(2)},
                    \dots,\widetilde{G}_{i-1}^{(S)}\}$ according to weights 
                        $\qty(\prod_{l=1}^{i-1} w_l^{(j)})^\alpha$.
                \item Split off a new district from $\widetilde{G}_{i-1}$
                    through one iteration of the splitting procedure
                    (Algorithm~\ref{algo:split}), creating a new plan
                    $(G_i,\widetilde{G}_i)$.
                \item If the newly sampled plan $(G_i,\widetilde{G}_i)$
                    satisfies $\pop(V_i)\in [P_i^-,P_i^+]$, save it;
                    otherwise, reject it.
            \end{enumerate}
        \item Calculate weights for each of the new plans \(
            w_i^{(j)} \ = \ \frac{\tau(G_i^{(j)})^{\rho-1}}{
                |\C(G_i^{(j)},\widetilde{G}_i^{(j)})|}.
                \)
        \end{enumerate}
    \item Calculate final weights 
    \vspace{-8pt}
    \begin{equation}
      w^{(j)} \ = \ \exp\{-J(\xi^{(j)})\} \psi([\xi^{(j)}])^{-1}
            \left(\prod_{i=1}^{n-2}w_i^{(j)}\right)^{(1-\alpha)}
            w_{n-1}^{(j)}\qty(\tau(G_{n}^{(j)}))^{\rho-1}.
    \end{equation}
    \item Output the $S$ final plans $\{\xi^{(j)}\}_{j=1}^{S}$, where
        $\xi^{(j)}=(G_1^{(j)},\ldots,G_{n-1}^{(j)},G_n^{(j)})$,
        and the final weights $\{w^{(j)}\}_{j=1}^{S}$.
    \end{enumerate} 
\end{algorithm}

From the output of Algorithm~\ref{algo:smc}, one last resampling of $S$ plans 
using the final weights can be performed to generate a final sample. 
Alternatively, the weights can be used directly to estimate the expectation of 
some statistics of interest under the target distribution, i.e., 
$H=\E_\pi(h(\xi))$, using the self-normalized importance sampling
estimate $\widehat{H} = \sum_{j=1}^S h(\xi^{(j)})w^{(j)}/ \sum_{j=1}^S w^{(j)}$.

The sampled plans are not completely independent, because the weights in
each step must be normalized before resampling, and because the
resampling itself introduces some dependence.  Precisely quantifying the
amount of dependence is difficult.
However, as we demonstrate in Section~\ref{sec:validation}, the dependence
is not large enough to cause measurable bias in summary statistics of interest.

It is difficult to precisely characterize
the computational complexity of the entire SMC algorithm since the
rejection sampling introduces a random component, which depends on
the difficulty of sampling a new district within the population bounds.
This random complexity is also shared by existing MCMC approaches,
which must redraw proposals if they are invalid.
Appendix~\ref{app:implementation} analyzes the complexity of generating each 
sample without accounting for the rejection step.
For both SMC and MCMC samplers, the total sampling time increases linearly in the sample size.
In contrast with MCMC approaches, however, step (b)(1) of Algorithm~\ref{algo:smc}
can be embarrassingly parallelized, since each resample-split-check requires 
interaction only with the previous set of samples.

The weights in the proposed algorithm are chosen to match existing
general SMC algorithms with partial rejection control. These existing
algorithms provide guarantees as to the convergence of the samples
to the target distribution. One such result, which will suffice for
our purposes, is the following central limit theorem.
\begin{restatable}{prop}{propvalid} \spacingset{1} \label{prop:valid}
    Let $\pi_S=\sum_{j=1}^S w^{(j)} \delta_{[\xi^{(j)}]}$ be the 
    weighted particle approximation generated by Algorithm~\ref{algo:smc}.
    Then for all measurable $h$ on unlabeled plans, as $S\to\infty$, \[
        \sqrt{S} \qty(\E_{\pi_S}[h([\xi])] - \E_{\pi}[h([\xi])])
        \cvd \Norm(0, V_\text{SMC}(h)),
    \] for some asymptotic variance $V_\text{SMC}(h)$.
\end{restatable}
A proof is given in Appendix~\ref{app:proofs}. This central limit theorem 
implies consistency (in $S$) of any quantity derived from the weighted samples. 
However, since this convergence is in probability (w.r.t. the algorithm's 
sampling probability), the proposition does not establish that $\pi_S\cvd \pi$ 
almost surely.
While the almost sure convergence result exists for a standard SMC algorithm
\citep{del2006sequential}, we do not know of an extension to the case 
of partial rejection control.

In some cases, the constraints incorporated into $J(\xi)$ admit a
natural decomposition to the district level as
$\prod_{i=1}^n J'(G_i)$---for example, a preference for districts
which split as few counties as possible, or against districts which
would pair off incumbents.  In these cases, an extra term of
$\exp\{-J'(G_i^{(j)})\}$ can be added to the weights $w_i^{(j)}$ in each
stage, and the same term can be dropped from the final weights
$w^{(j)}$. This can be particularly useful for more stringent
constraints; incorporating $J'$ in each stage allows the importance
resampling to ``steer" the set of redistricting plans towards those
which are preferred by the constraints.
This same idea can be used to amortize the contribution of $\psi([\xi])$ to
the final weights over the preceding SMC iterations; details can be found
in Appendix~\ref{app:implementation}.

\subsection{Practical Implementation and Use}
\label{subsec:details}

The SMC algorithm described above allows for an accurate characterization of 
the target distribution given in Equation~\eqref{eq:tgt} as the sample size goes
to infinity. In practice, of course, we must use a finite number of samples.
Additionally, the algorithm relies on choosing $k_i$, which can be challenging since 
$K_i$ is typically unknown, and on calculating $\psi([\xi])$, which
has no closed-from expression.

This section discusses these practical implementation challenges, as well as 
selection of the parameter $\alpha$.   We also describe the use of diagnostics 
(Section \ref{subsubsec:diag}) that can help identify when the 
algorithm is or is not performing well, and therefore if more samples or 
different constraints are required.
Further details about implementation may be found in Appendix~\ref{app:implementation}.

\subsubsection{Choosing \texorpdfstring{$k_i$}{k}} \label{subsubsec:k}

The accuracy of the algorithm is theoretically guaranteed only when
the number of edges considered for removal at each stage is at least
the maximum number of edges across all graphs which induce districts
$G_i$ with $\dev(G_i)\le D$, i.e., $k_i\ge K_i$.  Unfortunately, $K_i$
is generally unknown in practice.
We could conservatively set $k_i=|\tilde V_{i-1}|-1$, 
the number of edges in each spanning tree
but such a choice results in a prohibitively inefficient algorithm---the random
edge selected for removal will with high probability induce an invalid partition,
leading to a rejection of the entire map. 
In practice we estimate $k_i$ before each SMC stage by generating random spanning 
trees, as we detail in Appendix~\ref{app:implementation}. Theoretical support for
this estimate is given by Proposition \ref{prop:qksimp}, which bounds the 
inaccuracy of the approximation with a user-selectable parameter.

\subsubsection{Calculating \texorpdfstring{$\psi([\xi])$}{p([x])}} \label{subsubsec:psi}

The number of sequentially valid labelings $\psi([\xi])$ corresponding to an 
unlabeled plan $[\xi]$ can vary significantly across unlabeled plans, and
has no closed-form expression.
Unfortunately, $\psi([\xi])$ grows rapidly in $n$, but not nearly as fast as 
$n!$, the total number of labelings per unlabeled plan, which makes both direct 
and approximate calculation methods challenging.
We adopt a hybrid approach in practice. When $n\le 13$, we directly compute $\psi([\xi])$
with a recursive divide-and-conquer algorithm. When $n>13$, we can estimate
$\psi([\xi])$ to arbitrary precision using a particular importance sampling 
scheme. Both of these approaches, which are detailed in Appendix~\ref{app:implementation}, 
do not increase the runtime of the SMC algorithm appreciably.

\subsubsection{Choosing \texorpdfstring{$\alpha$}{a}}

As noted above, as long as $\alpha\in(0,1]$ the SMC algorithm is asymptotically 
valid. Larger values are more aggressive in downweighting unlikely plans 
(those which are over-represented in $q$ versus $\pi$), 
which may lead to less diversity in the final sample, while smaller
values of $\alpha$ are less aggressive, which can result in more variable final
weights and more wasted samples. 
\citet{liu2001} recommend a default choice of $\alpha=0.5$, which we have found
appropriate, though in general we have not found the algorithm's performance to 
be very sensitive to $\alpha$ within the range $0.5\le \alpha\le 0.9$.

\subsubsection{Diagnostics} \label{subsubsec:diag}

As with any complex sampler, application of the SMC algorithm in practice can be greatly 
facilitated by a set of diagnostic measures.  Diagnostics can never prove that
an algorithm is working correctly, but they can identify situations when the
algorithm fails and shed light on why.  We discuss several useful
diagnostics below, all of which are implemented in our open-source software.

Our primary recommendation is to perform multiple independent runs of the SMC 
algorithm, which enables both checking for nonconvergence to the target distribution
as well as the calculation of standard errors for quantities of interest.
For convergence, we adopt the Gelman-Rubin $\hat R$ statistic 
\citep{gelman1992inference}, which compares between-run variation to within-run variation.
If the latter is an appreciable fraction of the former, then independent runs
produce different results, and the algorithm has not converged.
For our purposes, we will consider an algorithm not to have converged if $\hat R$
values for statistics of interest exceed 1.05, though higher or lower thresholds 
may be appropriate for given computational budgets and other applied considerations.
If the algorithm has not converged for a particular $S$, analysts should run the 
algorithm again with a larger $S$, until $\hat R$ and other diagnostics discussed 
below indicate that the results are trustworthy.
At this point, the samples from the multiple runs can be combined to 
increase the precision of the estimates \citep[Chapter~11]{bda3}.

We use the rank-normalized and folded $\hat R$ of \citet{vehtari2019rank}, which
makes the statistic more robust to heavy tails and more sensitive to discrepancies 
in scale, not just location. For our purposes, $\hat R$ is advantageous, since 
it is computed on summary statistics and thus provides a measure of convergence
tailored to practical quantities of interest. Additionally, it is unitless, and
so can be compared across runs and indeed across algorithms, as we will do in
Section~\ref{sec:validation}.

Multiple runs can also be easily used to compute standard errors of expectations 
taken relative to the sampled plans.  Recently, \citet{lee2018variance} and
\citet{olsson2019numerically}, among others, have developed a method to estimate
SMC variance by using information on the ancestry of each sampled particle 
(here, plan). While we have implemented this method in our open-source software
and found it to be correct on average, the resulting estimates tend to
be noisier than those obtained from a simple multiple-run standard error.

Other diagnostics are useful for assessing the overall quality of the sample 
and pinpointing where issues arise when the algorithm is not working well.
One measure of quality is \textit{sample diversity}, or how different the 
samples are from each other, which we measure using the variation of
information metric shown in Equation~\eqref{eqn:varinfo}.  Similar
plans will have a variation of information near zero, while plans
which are extremely different will have a variation of information
closer to 1. A non-diverse sample will have many copies of similar or identical 
plans, which tends to increase sampling error and reduces the interpretability 
of the generated samples.

When sampling problems arise, a closer look at each iteration of 
Algorithm~\ref{algo:smc} can help identify the cause. A low acceptance rate
in step (b)(1)(3), or weights which are highly variable or have a heavy right 
tail in step (b)(2), can lower the sampling efficiency and diversity. Examining
the variance of the log-weights at each iteration can be useful. Variances
which increase consistently across iterations can be a sign to use a higher 
$\alpha$, which will sample more aggressively earlier on to avoid a build-up of 
variable weights.
We have also found that the number of unique plans from the previous iteration
which survive to the subsequent iteration is a highly useful indicator of sampling 
problems, since it captures inefficiencies introduced by both the rejection and 
resampling steps. If the number of unique plans is substantially below what would
be expected from a uniform sample with replacement from the same population, then
some kind of bottleneck or other sampling problem is generally present.

All of the diagnostics presented here are sample statistics, in that they vary 
across runs of the random SMC algorithm.
If the number of independent parallel runs were increased to infinity, 
these diagnostics would converge to a ``population value" for the particular 
sampling problem and choice of algorithm parameters.
But with a finite number of runs, there will be variation around this population
value, which can lead to diagnostic values that are too optimistic or pessimistic.
No diagnostic is foolproof.
However, taken together, the set of diagnostics presented here is designed to catch insufficient sample sizes and too-strong constraints with high probability.
Indeed, these diagnostics formalize and extend the so-called ``multi-start heuristic" 
advocated by some MCMC redistricting practitioners \citep{cannon2022spanning}.

\subsection{Incorporating Administrative Boundary Constraints}
\label{subsec:adminboundary}

Another common requirement for redistricting plans is that districts
``to the greatest extent possible" follow existing administrative
boundaries such as county and municipality lines.\footnote{If a
  redistricting plan must always respect these boundaries, we can
  simply treat the administrative units as the nodes of the original
  graph to be partitioned.}  In theory, this constraint can be
formulated using a $J$ function which penalizes maps for every county
line crossed by a district.  In practice, however, we can more
efficiently generate desired maps by directly incorporating this
constraint into our sampling algorithm.

Fortunately, with a small modification to the proposed algorithm, we
can sample redistricting plans proportional to a similar target
distribution but with the additional constraint that the number of
administrative splits not exceed $n-1$. 
The hard constraint of $n-1$ splits cannot be adjusted upwards or downwards, 
unfortunately, but a preference for even fewer 
administrative splits may be incorporated through the $J$ function.

Let $A$ be the set of administrative units, such as counties. We can
relate these units to the nodes by way of a labeling function
$\eta:V\to A$ that assigns each node to its corresponding unit. 
This function induces an equivalence relation $\sim_\eta$ on nodes, where
$v\sim_\eta u$ for nodes $v$ and $u$ iff $\eta(v)=\eta(u)$.  If we
quotient $G$ by this relation, we obtain the administrative-level
multigraph $G\,/\sim_\eta$, where each vertex is an administrative
unit and every edge corresponds to an edge in $G$ which connects two
nodes in different administrative units.  We can write the number of
administrative splits as
\begin{equation*}
    \spl(\xi)=\qty(\sum_{a\in A}\sum_{i=1}^n 
        C(\eta^{-1}(a)\cap\xi^{-1}(i))) -|A|,
\end{equation*}
where $C(\cdot)$ counts the number of connected components in the subgraph 
$\eta^{-1}(a)\cap\xi^{-1}(i)$.

\begin{figure}[!t]
  \centering \spacingset{1}
  \begin{subfigure}[t]{0.24\textwidth}
    \centering
    \includegraphics[width=1.4in]{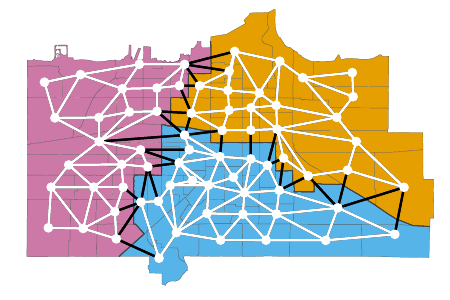}	
    \caption{\raggedright\small The Erie graph. Edges that cross from one administrative unit
      to another are colored black.}
  \end{subfigure}
  \begin{subfigure}[t]{0.24\textwidth}
    \centering
    \includegraphics[width=1.4in]{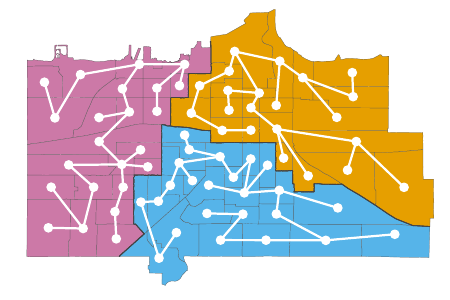}	
    \caption{\raggedright\small Spanning trees drawn on each administrative unit.}
  \end{subfigure}
  \begin{subfigure}[t]{0.24\textwidth}
    \centering
    \includegraphics[width=1.4in]{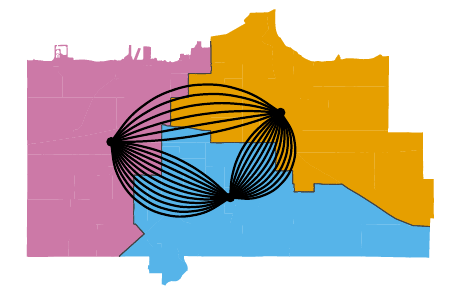}	
    \caption{\raggedright\small The quotient multigraph. Edges connecting boundary
    units in the original graphs become edges in the multigraph.}
  \end{subfigure}
  \begin{subfigure}[t]{0.24\textwidth}
    \centering
    \includegraphics[width=1.4in]{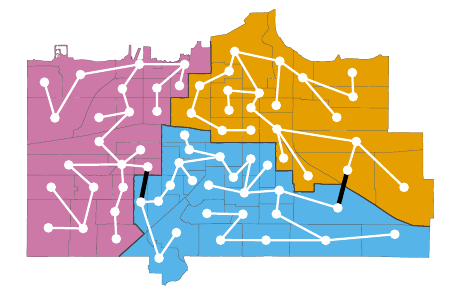}	
    \caption{\raggedright\small The final spanning tree. A spanning tree on the quotient
      multigraph connects the trees on each administrative unit.}
  \end{subfigure}
  \caption{The two-step spanning tree sampling procedure applied to the city
    of Erie, Pennsylvania, with three arbitrary administrative units indicated
    by the colored sections of each map.}
  \label{fig:hier}
\end{figure}

To implement this constraint, we draw the spanning trees in step~(a)
of Algorithm~\ref{algo:smc} in two substeps such that we sample from a specific
subset of all spanning trees. First, we use Wilson's algorithm to draw
a spanning tree on each administrative unit $a\in A$, and then we
connect these spanning trees to each other by drawing a spanning tree
on the quotient multigraph $\widetilde G_i\,/\sim_\eta$.
Figure~\ref{fig:hier} illustrates this process.  This approach is
similar to the independently-developed multi-scale merge-split
algorithm of \citet{autry2020}.

Drawing the spanning trees in two steps limits the trees used to those
which, when restricted to the nodes $\eta^{-1}(a)$ in any
administrative unit $a$, are still spanning trees. The importance of
this restriction is that cutting any edge in such a tree will either
split the map exactly along administrative boundaries (if the edge is
on the quotient multigraph) or split one administrative unit in two
and preserve administrative boundaries everywhere else. Since the
algorithm has $n-1$ stages, this limits the support of the sampling
distribution to maps with no more than $n-1$ administrative splits.

The two-step construction makes clear that the total number of
such spanning trees is given by
\begin{equation} \label{eq:hier-st}
    \tau_\eta(\widetilde G_i) \ = \ \tau(\widetilde G_i\,/\sim_\eta) 
    \prod_{a\in A} \tau(\widetilde G_i\cap \eta^{-1}(a)),
\end{equation} where $\widetilde G_i\cap \eta^{-1}(a)$ denotes the subgraph of
$\widetilde G_i$ which lies in unit $a$, and we take $\tau(\empty)=1$. 
Replacing $\tau$ with $\tau_\eta$ in the expression for the weights $w_i^{(j)}$ 
and $w^{(j)}$ then gives the modified algorithm that asymptotically samples from
\begin{equation}
    \label{eq:hier-tgt}
    \pi_\eta([\xi]) \ \propto \ \exp\{-J(\xi)\}\tau_\eta(\xi)^{\rho}
    \ind_{\{\xi\text{ connected}\}}\ind_{\{\dev(\xi)\le D\}}
    \ind_{\{\spl(\xi)\le n-1\}}.
\end{equation}

This idea can in fact be extended to arbitrary levels of nested
administrative hierarchy.  We can, for example, limit not only the number of
split counties but also the number of split cities and Census tracts to
$n-1$ each, since tracts are nested within cities, which are nested within
counties. 
To do so, we begin by drawing spanning trees using Wilson's algorithm on the 
smallest administrative units.
We then connect spanning trees into larger and larger trees by drawing
spanning trees on the quotient graphs of each higher administrative level. 
(In this approach, cities which cross county boundaries must be split
in two, and regions outside of cities must be treated as their own pseudo-city.) 
Even with multiple levels of administrative hierarchy, the
calculation of the number of spanning trees is still straightforward,
by analogy to Equation~\eqref{eq:hier-st}.

\section{An Empirical Validation Study} \label{sec:validation}

Although the proposed algorithm has desirable theoretical properties,
it is important to empirically assess its performance
\citep{fifield2020enum}. We examine whether or not the proposed
algorithm can produce a sample of redistricting maps that is actually
representative of a target distribution, and how the diagnostics of 
Section~\ref{subsubsec:diag} correlate with sampling accuracy.

\begin{figure}[!t]
  \centering \spacingset{1}
  \includegraphics[width=6.875in]{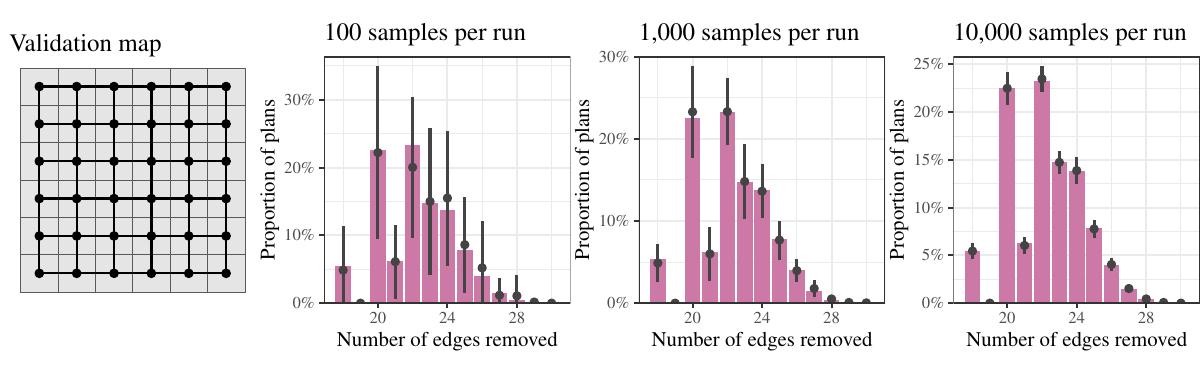}	
  \caption{The 6-by-6 map used in validation (left). 
    The remaining panels compare the distribution of the number of 
    removed edges in the enumerated distribution (purple bars) and the
    sampled distribution (grey) over a number of sample sizes $S$.
    The dots are the mean histogram estimate across the 24 independent
    runs, and represent the algorithm's bias.
    The vertical lines are 90\% confidence intervals for the histogram estimates,
    and represent the sampling variability.
    They are calculated from the standard deviation of the estimates 
    across the runs.}
  \label{fig:valid6x6}
\end{figure}

Our validation setting is the 6-by-6 grid map shown in Figure~\ref{fig:valid6x6}.
This map is small enough to obtain all possible redistricting plans with six
contiguous districts using an efficient enumeration algorithm \citep[see, e.g.,][]{fifield2020enum}.
Each precinct in the grid has an equal population. 
There are over 356 billion unlabeled plans with six districts, 
451,206 of which have districts with exactly balanced populations.\footnote{
The enumerated plans are publicly available at 
\url{https://github.com/gerrymandr/trees/tree/main/enumerations}.}
We demonstrate that the proposed algorithm can efficiently approximate a 
realistic target distribution on this set of balanced plans. 
Appendix~\ref{app:fl50} contains an additional validation study with four 
districts using a 50-precinct map taken from the state of Florida, where the 
population of each precinct varies.

We set the target distribution by choosing $\rho=1$, which as discussed above
will generate compact districts. Since this value of $\rho$ makes the 
proposal and sampling distributions as close as possible, the SMC algorithm 
will operate at peak efficiency. The validation study in 
Appendix~\ref{app:fl50} explores a wider range of values of $\rho$.
As we have noted in Section \ref{sec:goal}, values of $\rho$ other than 1 
may not be computationally feasible for larger redistricting problems with
many more precincts.

We run Algorithm~\ref{algo:smc} on a logarithmically-spaced range of 
sample sizes $S$ ranging from 10 to 10,000.
While $S=10$ is far too small to use in any real analysis, we include it here 
for illustrative purposes to demonstrate the high variability and potential bias
of the algorithm when $S$ is too small.
At each sample size, we perform 24 independent runs of the algorithm, 
each of size $S$, which allows for accurate calculation of standard errors 
and $\hat R$ for summary statistics.

To validate the algorithm, we compare the distribution of a summary 
statistics from each sample to the true distribution obtained by reweighting
the enumerated plans according to the target measure, which is proportional to
$\tau(\xi)$.
The summary statistic used here is the number of edges that must be removed 
to form each set of districts, $\rem(\xi)\cdot|E(G)|$.
As discussed in Section~\ref{subsec:comp}, this statistic is highly correlated
with the target $\tau(\xi)$, and so is a good test of the algorithm.

\begin{figure}[!t]
  \centering \spacingset{1}
  \includegraphics[width=6.875in]{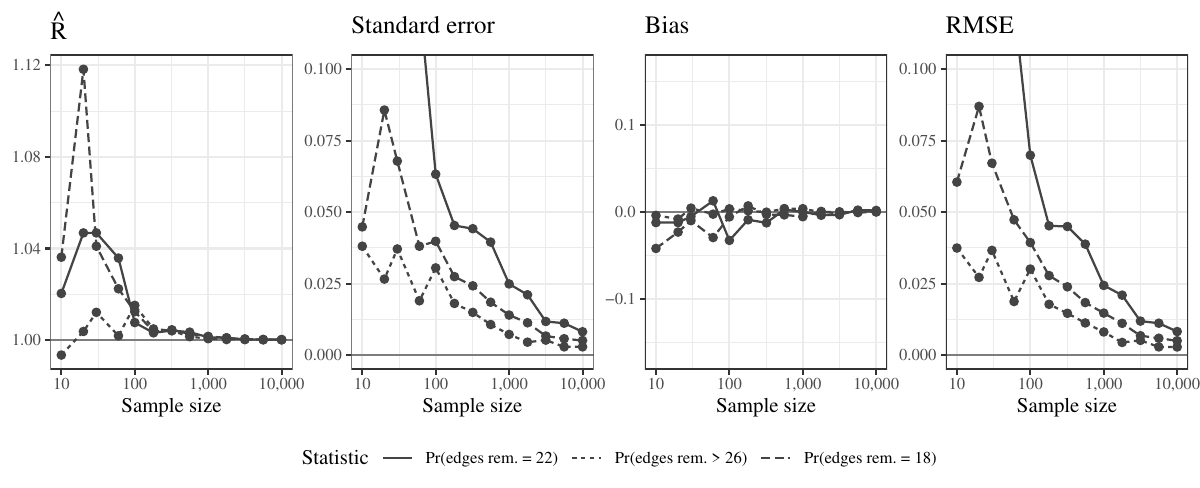}
  \caption{Convergence diagnostic $\hat R$ (on a log scale), standard
    errors, bias, and RMSE for the same three compactness summary statistics, 
    calculated across 24 independent runs of the SMC algorithms.
    Values are plotted versus sample size $S$ per run.} 
  \label{fig:diagn6x6}
\end{figure}

These comparisons are summarized in the right three panels of 
Figure~\ref{fig:valid6x6}, which show the true distribution of the number of 
edges removed in purple, with the estimated histogram overlaid as solid points, 
for a range of sample sizes.
Each histogram estimate produced by the SMC algorithm is accompanied by a 90\%
confidence interval produced using the standard errors calculated across the
24 independent runs (i.e., the standard deviations of the estimates
for each run). 
With as few as $S=100$ samples, the SMC estimates are already close to the true
distribution.
In fact, if we tested the null hypothesis that each estimate is centered at 
the true value, we would only reject for one of the twelve nonzero 
true values at the 10\% level, within the range of what would be expected by 
chance given the sampling variability.
As the sample size increases, the variance of the SMC estimates decreases,
until by $S=10,000$ samples it is mostly negligible.
Overall, the agreement between the sampling and target distributions is 
excellent, indicating that the SMC algorithm is able to accurately sample from 
the target distribution in this setting.

We next study more quantitatively the quality of the SMC sample as the 
sample size increases, and in particular how well the diagnostics recommended
in Section~\ref{subsubsec:diag} can be used as a proxy for this quality.
To do so, we focus on three particular estimands that can be obtained from
the distribution of edges removed: the probability of removing exactly 22 edges
(the target median; true probability 0.2328), 
the probability of removing exactly 18 edges (true value 0.0542),
and the probability of removing more than 26 edges (true value 0.0204).
The latter two statistics are especially useful as tests of the algorithm's 
sampling accuracy, since they require estimating small tail probabilities which
may correspond to very few redistricting plans.
On the lower tail, there are just 2 possible unlabeled plans which remove exactly 18 
edges, though together they comprise over 5\% of the target distribution.
The upper tail probability corresponds to the $p$-value a researcher would use
for testing whether an enacted plan with 26 removed edges was significantly
less compact than would be expected under the target distribution.

For each sample size from the SMC algorithm, we calculate the $\hat R$ diagnostic,
and standard errors for these three summary statistics.
Note that neither of these calculations requires access to the enumerated 
ground truth.
We also compute the bias and the root mean-square error (RMSE) for the estimates,
comparing to the true values from the enumerated target distribution.
Ideally, the always-computable $\hat R$ and standard errors will well represent
the generally uncomputable bias and RMSE.

Figure~\ref{fig:diagn6x6} summarizes the results of this experiment.
As expected, as the SMC sample size $S$ increases, $\hat R$, the standard 
errors, and the relative RMSE also decrease.
The $\hat R$ values fall below our recommended cutoff of 1.05 by $S=60$, 
indicating that for samples this large and greater, each independent run is 
producing similar estimates and the SMC algorithm has likely converged to 
the target distribution.
Indeed, the estimates' bias is closely centered around zero in this range.
However, standard errors are significantly larger for the smaller sample sizes.
This tracks the pattern observed in the validation above: at small sample
sizes, the SMC estimates were on average unbiased for the target 
distribution, but had significant sampling variability.

Importantly, the calculated standard errors track the actual RMSE extremely 
closely, demonstrating their value as an observable proxy for this measure of 
sampling accuracy.
In other words, with low $\hat R$ indicating likely algorithmic convergence 
(and therefore unbiasedness), the overall algorithmic error is driven
almost entirely by the sampling variability, which is measurable with
the standard errors.

\section{Analysis of the 2011 Pennsylvania Redistricting} \label{sec:pa-study}

As discussed in Section~\ref{sec:pa-overview}, in the process of
determining a remedial redistricting plan to replace the 2011 General
Assembly map, the Pennsylvania Supreme Court received submissions from
seven parties. In this section, we compare four of these maps to both
the original 2011 plan and the remedial plan ultimately adopted by the
court. We study the governor's plan and the House Democrats' plan; the
petitioner's plan (specifically, their ``Map A"), which was selected
from an ensemble of 500 plans used as part of the litigation; and the
respondent's plan, which was drawn by Republican officials. 

\subsection{The Setup}

To evaluate these six plans, we drew reference maps from the
target distribution given in Equation~\eqref{eq:hier-tgt} by using the
proposed algorithm along with the modifications presented in
Section~\ref{subsec:adminboundary} to cap the number of split
counties at 17 (out of a total of 67), in line with the court's
mandate.  We set $\rho=1$ to put most of the sample's mass on
compact districts, and enforced $\dev(\xi)\le 0.001$ to reflect the
``one person, one vote" requirement. 

This population constraint translates to a tolerance of around 700
people, in a state where the median precinct has 1,121 people.  Like
most research on redistricting, we use precincts because they
represent the smallest geographical units for which election results
are available.  To draw from a stricter population constraint we would
need to use the 421,545 Census blocks in Pennsylvania rather than the
9,256 precincts, which would significantly increase the computational
burden.

\subsection{Comparison with an Analogous MCMC Algorithm} \label{subsec:compare}
  

We first compare the computational efficiency of the SMC algorithm
with a merge-split-type MCMC algorithm \citep{carter2019,deford2019}.
The merge-split algorithm here uses a spanning tree-based proposal
identical to the splitting procedure described in
Algorithm~\ref{algo:split}: it merges adjacent districts, draws a
spanning tree on the merged district, and splits it to ensure the
population constraint is met.
A Metropolis-Hastings correction is used, analogous to the SMC reweighting step.
In fact, the implementation (contained in our open-source \texttt{redist} package) 
shares much of the same code base, which allows for performance comparisons
that better reflect the different algorithmic strategies, rather than different
algorithmic implementations.\footnote{A particularly performant implementation
of a similar algorithm \citep[Reversible ReCom,][]{cannon2022spanning} 
may be found at \url{https://github.com/pjrule/frcw.rs}}

The MCMC chains are each initialized with a random SMC sample, 
and run for 500 warm-up iterations before samples are recorded.
We let the sample size vary from 100 to 4,000 for the SMC algorithm and 
from 100 to 8,000 for the MCMC algorithm. For each sample size, we
perform four independent runs of each algorithm.
The average acceptance rate for the MCMC algorithm across the four chains 
was 88.7\%.

\begin{figure}[!t]
	\centering \spacingset{1}
    \includegraphics[width=6.75in]{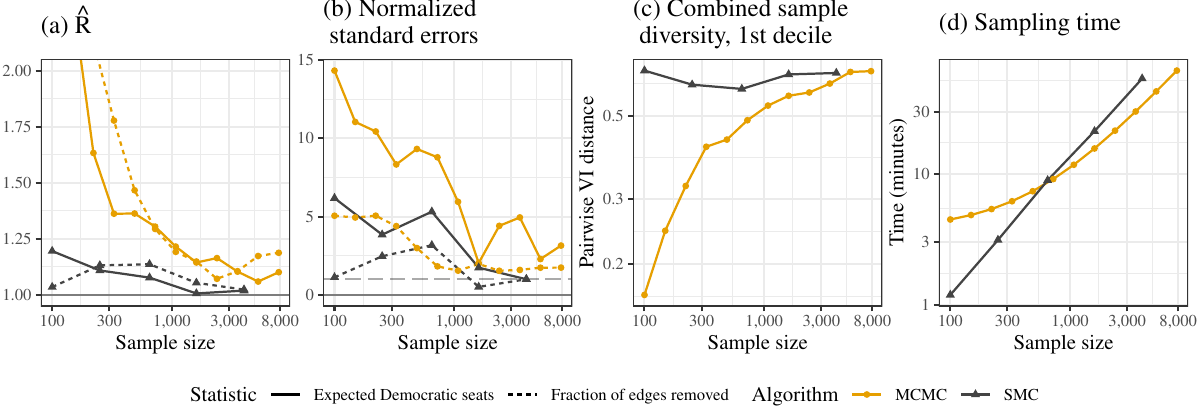}	
    \caption{A comparison of sampling efficiency for SMC and MCMC algorithms 
    using the same transition kernel.
    (a) Convergence diagnostic $\hat R$ for two summary statistics plotted 
    against the sample size.
    (b) Standard errors for two statistics, normalized by the standard error
    for the SMC algorithm at 4,000 samples (to put the two statistics on the
    same scale), by sample size.
    (c) First decile of the pairwise variation of 
    information distance (a measure of sample diversity) by sample size.
    Distance measures were computed with the combined sample of 
    four independent runs. 
    (d) Sampling time by sample size.}
    \label{fig:pa-compare}
\end{figure}

Figure~\ref{fig:pa-compare}(a) shows the $\hat R$ values of two summary 
statistics (Democratic seats and compactness) by sample size $S$ per run. 
By this convergence heuristic, the SMC algorithm converges more quickly, 
with lower $\hat R$ values than the MCMC algorithm for a given sample size. 
According to the check that $\hat R\le 1.05$, the SMC algorithm has 
approximately converged with $S=1,600$, while the MCMC algorithm has 
still not converged after with than 8,000 post-warmup samples (over 7,000 unique 
accepted proposals), at which point the MCMC algorithm has been running 2.4 times 
longer than the SMC algorithm.

We find a similar story in examining the standard errors of the two statistics,
which are shown in Figure~\ref{fig:pa-compare}(b) after being normalized to
be on comparable scales.
With four independent runs per algorithm, standard error estimates themselves
are rather noisy, but the overall trend mirrors the findings for $\hat R$.
Even at large sample sizes, the MCMC standard errors are 1.5--5 times larger than
their SMC counterparts, although the MCMC standard errors are not meaningful without the underlying
chains having converged.

Figure~\ref{fig:pa-compare}(c) examines the sampling efficiency from the 
perspective of sample diversity. As discussed in Section~\ref{subsubsec:diag},
the pairwise variation of information (VI) distance provides a way to quantify how
similar the plans in each sample are to one another. If the distribution of these
pairwise VI distances has a long lower tail, then there are a significant number of
plans which are very similar. Figure~\ref{fig:pa-compare}(c) shows the first 
decile of the pairwise VI distances in the combined sample (i.e., pooled across 
all four runs) for each algorithm as the sample size increases. Because of the
sequential nature of the MCMC algorithms, the sample diversity doesn't match that
of the SMC algorithms until there are at least 3,000 samples. This is reflected
in the lag-1 autocorrelation of the MCMC sample, which was 0.982 for Democratic 
seats and 0.984 for compactness. 
Autocorrelation is of course not a direct measure of sampling efficiency, 
but this comparison serves to underscore the differences in the types of samples
generated by each algorithm for finite sample sizes.

Figure~\ref{fig:pa-compare}(d) shows how long each algorithm takes to run for
a given number of samples. Because of the fixed warm-up cost of the MCMC algorithm,
it is slower than the SMC algorithm for fewer than 600 samples. Asymptotically, 
both algorithms have runtime linear in the number of samples. But the 
per-sample cost of the MCMC algorithm is lower, since only two districts are 
redrawn, rather than the entire map.

\subsection{Compactness and County Splits}

While the results of the comparison above suggest that 1,600 SMC samples is 
enough for Pennsylvania, we perform the rest of our analysis with 4,000 samples
from one of the SMC runs.
Apart from the $\hat R$ diagnostic examined in detail above, the other SMC
diagnostics of Section~\ref{subsubsec:diag} all indicated no problems with the
generated sample.

\begin{figure}[!t]
  \centering \spacingset{1}
  \includegraphics[width=6.75in]{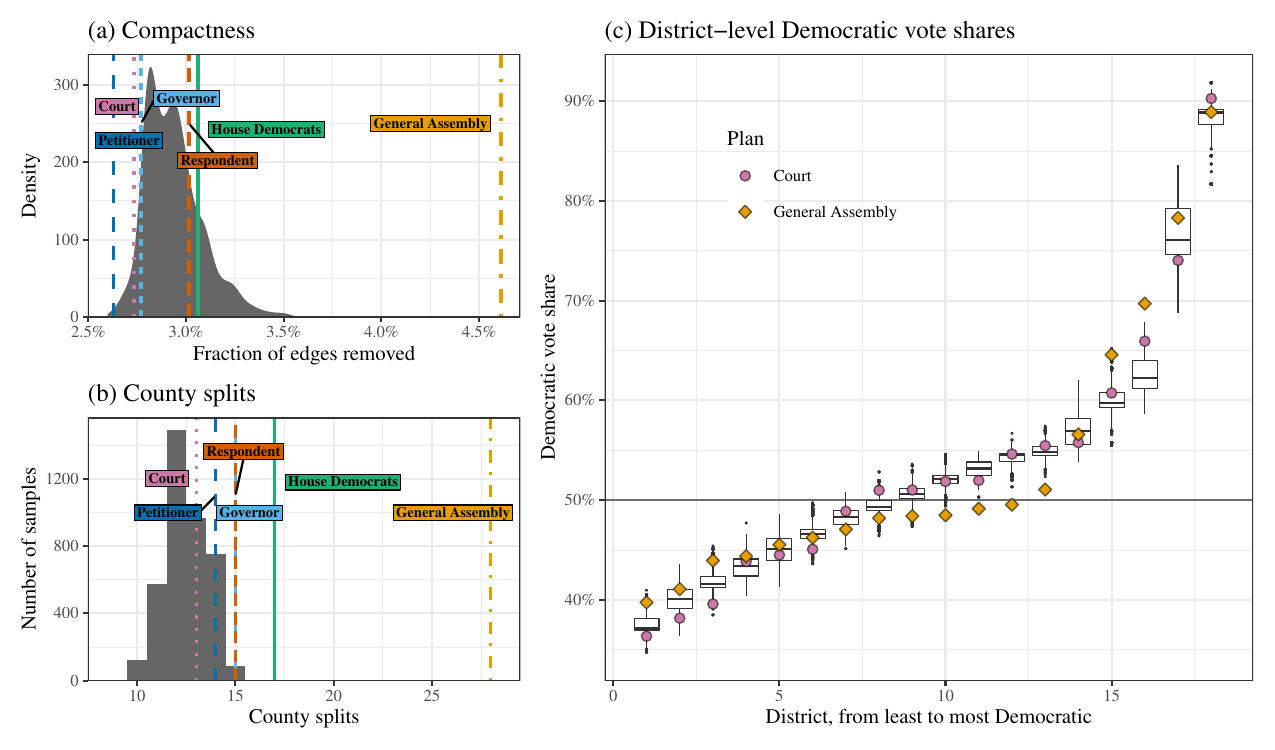}
  \caption{Summary statistics for the sampled and comparison plans:
  (a) compactness (smaller means  more compact) and (b) county splits. 
  Panel (c) shows the Democratic two-party vote share by 
  district, where within each plan districts are ordered by Democratic vote share.}
  \label{fig:pa-sum}
\end{figure}

Figures~\ref{fig:pa-sum}(a) and (b) show distribution of the fraction of
edges removed $\rem(\xi)$ and the number of county splits
$\spl(\xi)$ across the reference maps generated by our algorithm. 
The figure also shows these values for each of the six plans.
The 2011 General Assembly plan is a clear outlier for both statistics, 
being far less compact and splitting far more counties than any of the reference
plans and all of the remedial plans.  Among the remedial plans, the
petitioner's is almost an outlier in being \textit{too} compact, although this 
is perhaps not surprising---the map was generated by an algorithm
explicitly designed to optimize over criteria such as population
balance and compactness \citep{chen2017expert}. The other plans' compactness
falls within the range of the reference sample.

Figure~\ref{fig:pa-sum}(b) shows that all of the submitted remedial plans, as
well as the reference sample, followed the court's instructions to limit the
number of county splits. Yet around half of the reference maps split fewer 
than 13 counties, the minimum number of splits found in any of the remedial plans. 
This may be a result of the strict population constraint (all six plans were 
within 1 person of equal population across all districts), or a different 
prioritization between the various constraints imposed.

\subsection{Partisan Analysis}

While important, the outlier status of the General Assembly plan as
regards compactness and county splits is not sufficient to show that
it is a \textit{partisan} gerrymander.  To evaluate the partisan
implications of the six plans, we take a precinct-level baseline
voting pattern and aggregate it by district to explore hypothetical
election outcomes under the six plans and the reference maps.  The
baseline pattern is calculated by averaging the vote totals for the
three presidential elections and three gubernatorial elections that
were held in Pennsylvania from 2000 to 2010.\footnote{Data from
  Ansolabehere, S.  and Rodden, J. (2011). Pennsylvania Data
  Files. Available at \url{https://doi.org/10.7910/DVN/FJHHDS}.}
These election data were also used during litigation.  While being far
from a perfect way to create counterfactual election outcomes, this
simple averaging of statewide results is often used in academic
research and courts. The values of $\hat R$ for the mean and various 
quantiles of the district-level vote shares were all within standard 
convergence ranges.

Then, within each plan, we number the districts by their baseline
Democratic two-party vote share, so District~1 is the least Democratic
and District~18 the most. Figure~\ref{fig:pa-sum}(c), analogous to
Figure~7 in \citet{herschlag2017}, presents the distribution of the
Democratic two-party vote share for each of the districts across the
reference maps, and also shows the values for the General Assembly
plan (orange triangles) and the court's adopted plan (purple circles).
When compared to the reference maps and the court's plan, the General
Assembly plan tends to yield smaller Democratic vote share in
competitive districts ($p$-values of 0.0162, 0.0002, 0.0002, and
0.0002 for Districts 9--12 compared to the reference set) while giving
larger Democratic vote share in non-competitive districts. The effect
of this is to produce 6 Democratic seats, on average, under the
General Assembly plan, compared to a range of 8--12 under the
reference sample ($p$-value 0.0002) and 11 under the Court plan.  All
of these findings provide evidence that the General Assembly plan was
gerrymandered in favor of Republicans by packing Democratic voters
into non-competitive districts.

\section{Concluding Remarks} \label{sec:conclusion}

Redistricting sampling algorithms allow for the empirical evaluation
of a redistricting plan by generating alternative plans under a
certain set of constraints.  Researchers and policymakers can compute
various statistics from the redistricting plan of interest and
compare them with the corresponding statistics based on these
sampled plans.  Unfortunately, existing approaches often struggle when
applied to real-world problems, owing to the scale of the problems and
the number of constraints involved.

The SMC algorithm presented here can provably asymptotically sample from a 
specific target distribution.
Because of the constructive sampling approach, it does not face the problem of 
potential reducibility that MCMC algorithms have, and 
performs more efficiently in a realistic applied setting 
than a MCMC algorithm that uses an identical  proposal distribution.
The proposed method also incorporates, by design, the common redistricting 
constraints of population balance, geographic compactness, and 
minimizing administrative splits.
We expect these advantages of the SMC algorithm to
meaningfully improve the reliability of outlier analysis in real-world
redistricting cases. 

The proposed approach has some important limitations. Like other redistricting
sampling algorithms, if the target distribution strays too far from the proposal
distribution, the quality of the generated sample will suffer dramatically.
Close examination of diagnostic measures like $\hat R$, standard errors,
sample diversity, and per-iteration efficiency measures like weight variance 
and the number of surviving plans is required to provide confidence 
that the algorithm is sampling accurately and has converged for 
particular quantities of interest.
Our open-source software provides these diagnostics so that
analysts can use the proposed algorithm successfully.

One important drawback particular to the SMC algorithm arises in situations with dozens
or hundreds of separate districts.  Like any other SMC or particle resampling
scheme, as the number of iterations (districts) grows, eventually all of the
samples will share a common ancestor. While this is not a problem in many SMC 
applications (such as Bayesian inference), for redistricting, this means that
all of the sampled plans will share one or more districts that are completely 
identical. Any summary statistics which rely on these districts will have
very large standard errors and may not converge even with a large number of 
samples. 
While this problem will be readily identified with the diagnostics we propose
here, there is little the analyst can do to address the issue beyond increasing
the sample size and possibly combining many independent runs of the algorithm. 
One interesting avenue for future 
research would be to examine whether several applications of an MCMC kernel
at various points in the SMC algorithm could help refresh the sample and 
counteract the tendency to collapse to a single ancestor.

Future research should also explore the possibility of improving several
design choices in the algorithm to further increase its efficiency.
Wilson's algorithm, for instance, can be generalized to sample from
edge-weighted graphs.  Choosing weights appropriately could lead to
trees which induce maps that are more balanced or more compact.  And
the procedure for choosing edges to cut, while allowing for the
sampling probability to be calculated, introduces inefficiencies by
leading to many rejected maps.  Further improvements in either of
these areas should allow us to better sample and investigate
redistricting plans over large maps and with even more complex sets of
constraints.  

Finally, we believe that the application of these sampling algorithms
to real-world redistricting problems is essential.  We have applied
the SMC algorithm to the 2020 Congressional redistricting in all 50
states \citep{50stateSimulations}, and this allowed us to evaluate the
degree of partisan gerrymandering across states
\citep{gerrymandering2022}.  Such applications shed light on
substantive political science theories, and highlight new
methodological challenges.

\bigskip
\pdfbookmark[1]{References}{References}
\bibliography{citations}


\newpage

\appendix
\renewcommand\thefigure{\thesection.\arabic{figure}} \setcounter{figure}{0}

\section{Proofs of Propositions}
\label{app:proofs}

\splitprob*
\begin{proof}
    Any spanning tree can be decomposed into two other trees and an edge
    joining them.  Let $T\cup e\cup T'$ denote the spanning tree obtained
    by joining two other spanning trees, $T$ and $T'$, with an edge $e$.
    Then Equation~\eqref{eq:rhs-split-pr} can be written as \[
        q(G_i\mid\widetilde{G}_{i-1}) 
        = \sum_{\substack{T^{(1)}\in \mathcal{T}(G_i) \\ 
                           T^{(2)}\in \mathcal{T}(\widetilde{G}_i)}} 
            \sum_{e\in \C(T^{(1)}, T^{(2)})} 
            q(G_i\mid T^{(1)}\cup e\cup T^{(2)})\,
            \tau(\widetilde{G}_{i-1})^{-1}. 
    \] 
    Now, $q(G_i\mid T^{(1)}\cup e\cup T^{(2)})$ is determined by whether
    whether $e^*=e$, i.e., if $e$ is the edge selected to be cut. If $e$
    has $d_e$ in the top $k_i$ (if it induces one of the best $k_i$
    balanced splits), then it has a $1/k_i$ probability of being
    selected in step~(c) and cut. If $d_e$ is not in the top $k_i$, then
    this probability is zero.

    Everything written to this point holds regardless of whether 
    $G_i$ is a valid district (i.e., satisfies
    $\pop(V_i)\in[P_i^-,P_i^+]$). From here onwards we will restrict our
    attention to valid districts only. 
    Notice that the forward-looking bounds $P_i^-$ and
    $P_i^+$ are stricter than merely ensuring $\dev(G_i)\le D$. That is,
    conditional on $\pop(V_i)\in[P_i^-,P_i^+]$, we must also have
    $\dev(G_i) \le D$. 

    Therefore, if a sorted edge $e_j$ \textit{in any spanning tree}
    induces such a balanced partition, we must have $j \le K_i$, where
    as in the main text $K_i$ counts the maximum number of such edges
    across all possible spanning trees.  Thus, so long as we set $k_i\ge
    K_i$, we will have $d_e\le D$.

    Furthermore, across all spanning trees $T^{(1)}\in \mathcal{T}(G_i)$
    and $T^{(2)} \in \mathcal{T}(\widetilde{G}_i)$, and connecting edges
    $e\in E(T^{(1)},T^{(2)})$, the value of $d_e$ is constant, since
    removing $e$ induces the same districting. Combining these two facts,
    we have, conditional on satisfying the bounds $P_i^-$ and $P_i^+$,
    \begin{equation*}
        q(e^*=e\mid T^{(1)}\cup e\cup T^{(2)},\pop(V_i)\in[P_i^-,P_i^+])
      \ = \ k_i^{-1},
    \end{equation*}
    which does not depend on $T^{(1)}$, $T^{(2)}$, or $e$. We may
    therefore write the conditional sampling probability as
    \begin{align*}
        q(G_i\mid\widetilde{G}_{i-1}, \pop(V_i)\in[P_i^-,P_i^+])
        & \ = \ \sum_{\substack{T^{(1)}\in \mathcal{T}(G_i) \\ 
                           T^{(2)}\in \mathcal{T}(\widetilde{G}_i)}} 
            \sum_{e\in \C(T^{(1)}, T^{(2)})} 
            \frac{1}{k_i \tau(\widetilde{G}_{i-1})} \\
        &\ = \ \frac{\tau(G_i)\tau(\widetilde{G}_i)}{\tau(\widetilde{G}_{i-1})k_i} 
            |\C(G_i, \widetilde{G}_i)|,
            \numberthis \label{eq:simp-stage}
    \end{align*}
    where as in the main text we let $\C(G, H)$ represent the set of
    edges joining nodes in a subgraph $G$ to nodes in a subgraph $H$.
\end{proof}
    
\propvalid*
The proof proceeds by showing that the weights in
Algorithm~\ref{algo:smc} are of a form derived from an existing 
SMC algorithm with an established central limit theorem.
\begin{proof}
    We can associate our target measure $\pi([\xi])$ on unlabeled redistricting
    plans with a corresponding measure on labeled plans \[
        \tilde\pi(\xi) \dfeq \psi([\xi])^{-1}\pi([\xi]),
    \] so that the pushfoward measure obtained by mapping $\xi\mapsto[\xi]$ 
    recovers $\pi$. 
    
    Our SMC algorithm will operate on labeled plans, targeting $\tilde\pi$, so
    that the resulting plans, when considered as representatives of their
    corresponding unlabeled plans, will be representative of $\pi$ in the 
    sense given by the theorem statement.
    Recall that a labeled redistricting plan $\xi$ is just a tuple of graph
    partitions $(G_1, G_2, \dots, G_n)$. We begin by extending $\tilde\pi(\xi)$ 
    to a series of measures on partial plans, 
    \begin{align*}
        \tilde\pi_i(G_1,G_2,\dots,G_i) &:\propto
        \prod_{j=1}^i \frac{\tau(G_j)^\rho\tau(\widetilde G_j)}{
            \tau(\widetilde G_{j-1})} \ind_{\pop(V_j)\in[P_j^-,P_j^+]} \\
        &\propto \tau(\widetilde G_j) \prod_{j=1}^i \tau(G_j)^\rho
            \ind_{\pop(V_j)\in[P_j^-,P_j^+]}, 
    \end{align*}
    for $1\le i\le n-2$, and where we have simplified the telescoping
    product in the second equality.  Recall that the $\widetilde G_i$
    are determined completely by $G_1,G_2,\dots, G_i$. 

    For $i=n-1$, the above definition would yield \[
        \tilde\pi_{n-1}(G_1,G_2,\dots,G_i) \propto
        \tau(\widetilde G_{n-1}) \prod_{j=1}^{n-1} \tau(G_j)^\rho
            \ind_{\pop(V_j)\in[P_j^-,P_j^+]}, 
        = \tau(\xi) \tau(G_n)^{1-\rho} \ind_{\dev(\xi)\le D},
    \] which is close to but not quite the target measure. So we instead
    define $\pi_{n-1}\dfeq \tilde\pi$; i.e., we add the additional
    terms $\exp(-J(\xi))$ and $\psi([\xi])$ and adjust for $\tau(G_n)^{1-\rho}$. 

    With these partial-plan measures defined, notice that the
    incremental weight $w_i^{(j)}$ for partial plans with
    $1\le i\le n-2$ and $\pop(V_j)\in[P_j^-,P_j^+]$ may be written as 
    \begin{align*}
        w_i^{(j)} 
        &= \tau(G_i^{(j)})^{\rho-1}
            \frac{k_i}{|\C(G_i^{(j)},\widetilde{G}_i^{(j)})|} \\
        &= \frac{\tau(G_i^{(j)})^\rho\tau(\widetilde G_{i}^{(j)})}{
            \tau(\widetilde G_{i-1}^{(j)})}
            \qty(\frac{\tau(G_i^{(j)})\tau(\widetilde G_i^{(j)})}{
            \tau(\widetilde G_{i-1}^{(j)})}
            \frac{|\C(G_i^{(j)},\widetilde{G}_i^{(j)})|}{k_i})^{-1} \\
        &= \frac{\tilde\pi_i(G_1,\dots,G_i)}{\tilde\pi_{i-1}(G_1,\dots,G_{i-1})
            q(G_i\mid\widetilde{G}_{i-1}, \pop(V_i)\in[P_i^-,P_i^+])}.
            \numberthis\label{eq:inc-weight}
    \end{align*}
    For the final weighting at split $i=n-1$, the incremental weight
    (i.e., not including the residual previous weights
    $\qty(\prod_{i=1}^{n-2} w_i^{(j)})^{1-\alpha}$ given by step (c) of
    Algorithm~\ref{algo:smc} is 
    \begin{align*}
        \exp(-J(\xi^{(j)}))&w_{n-1}^{(j)}\psi([\xi])^{-1}\qty(\tau(G_{n-1}^{(j)}))^{\rho-1} \\
        &= \frac{\tilde\pi_{n-1}(G_1,\dots,G_{n-1})}{\tilde\pi_{n-2}(G_1,\dots,G_{n-2})
            q(G_{n-1}\mid\widetilde{G}_{n-2}, 
            \pop(V_{n-1})\in[P_{n-1}^-,P_{n-1}^+])},
    \end{align*}
    since this weight includes exactly the same additional terms as
    $\tilde\pi_{n-1}$ mentioned above. So in fact
    Equation~\eqref{eq:inc-weight} holds for all $1\le i\le n-1$.

    These incremental weights are precisely those of the SMC partial
    rejection control algorithm of \citet{peters2012sequential} (see
    also \citet{legland2005sequential}), with the weights set to zero
    for invalid samples and the partial rejection threshold set to the
    minimum possible nonzero weight. So after pushing forward to unlabeled plans,
    we gain immediately the theorem proved in that work (its Equation~6), viz., 
    that for all measurable $h$ on unlabeled plans and as $S\to\infty$, we have
    \[ \sqrt{S} \qty(\E_{\pi_S}[h([\xi])] - \E_{\pi}[h([\xi])]) \cvd
      \Norm(0, V_\text{SMC}(h)),
    \] for asymptotic variance $V_\text{SMC}(h)$ given by Equation~7 of
    the same work.
\end{proof}

\section{Additional Validation Example}
\label{app:fl50}

This section reports the results of another validation study applied to
a 50-precinct map taken from the state of Florida.
As in Section \ref{sec:validation}, we use the efficient enumeration
algorithm of \citet{fifield2020enum} to obtain all possible
redistricting maps with three and four contiguous districts, and use these
plans as a baseline to validate the proposed algorithm.
The left plot of Figure~\ref{fig:fl50-overview} below shows the validation map.

\begin{figure}[H]
    \centering \spacingset{1}
    \begin{subfigure}[c]{0.4\textwidth}
        \centering
        \includegraphics[width=2.7in]{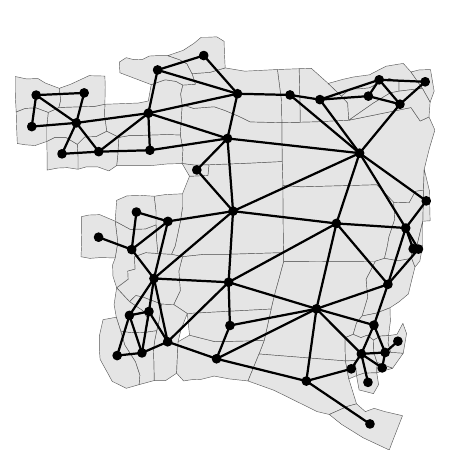}	
    \end{subfigure}
    \hspace{0.05\textwidth}
    \begin{subfigure}[c]{0.4\textwidth}
        \centering
        \includegraphics[width=2.8in]{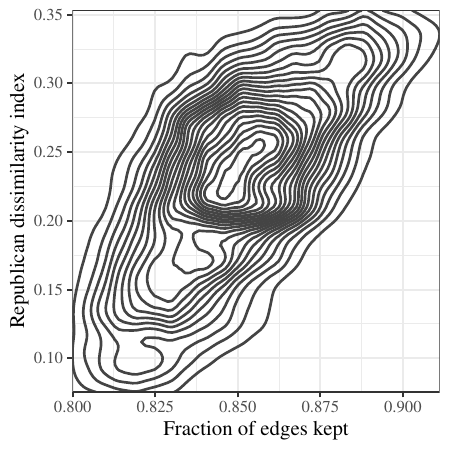}	
    \end{subfigure}
    \caption{The 50-precinct Florida map used for validation (left) and the 
    joint distribution of Republican dissimilarity and compactness on the map 
    over all partitions into four districts with $\dev(\xi)\le 0.10$ (right).}
    \label{fig:fl50-overview}
\end{figure}

There are 112,515,494 partitions of the map into four districts, 33,635 of which
have $\dev(\xi)\le 0.10$. We evaluate the accuracy of the proposed algorithm,
and compare it to the same MCMC algorithm used in Section~\ref{sec:pa-study},
across a range of target distributions, with $\rho$ running from 0.8 to 1.2
The right plot of Figure~\ref{fig:fl50-overview} shows the joint distribution 
of compactness and the summary statistic we use for the validation, the 
Republican dissimilarity index \citep{massey1988}.
With this validation map, Republican dissimilarity is reasonably sensitive to
the compactness of districts. This makes dissimilarity a good test statistic for
comparing distributions that differ primarily in their average compactness.

\begin{figure}[!t]
  \centering \spacingset{1}
  \includegraphics[width=6.5in]{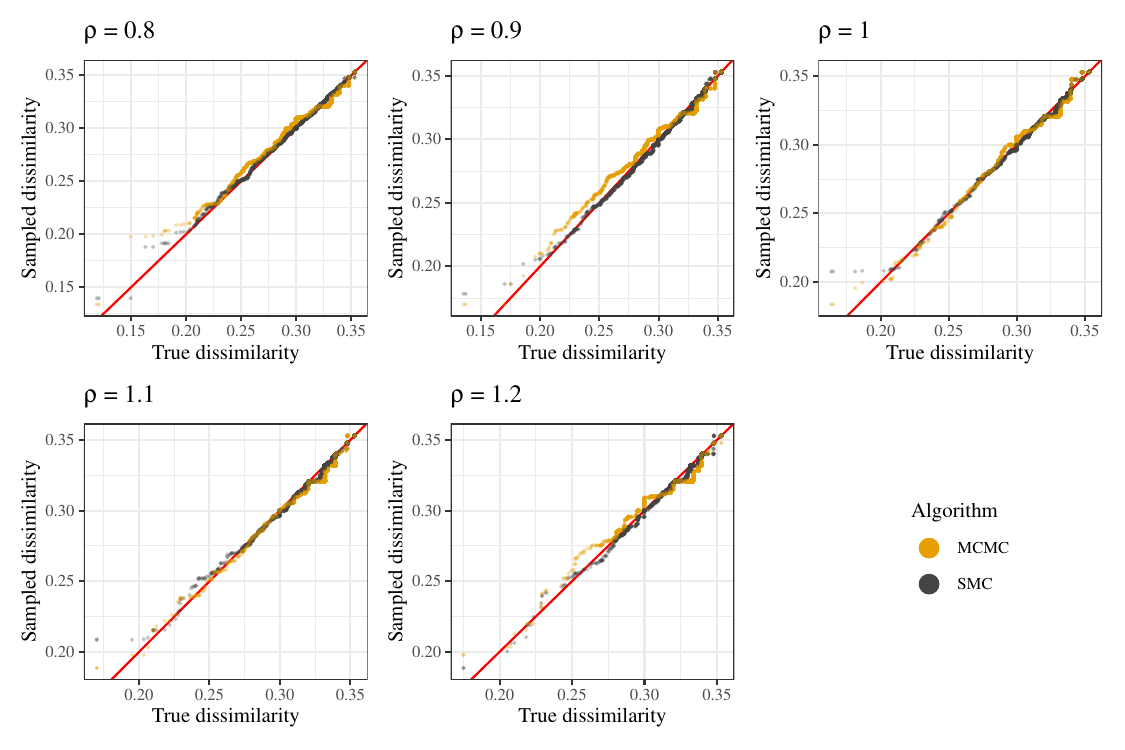}
  \caption{Quantile-quantile plots for 1,500 MCMC and SMC samples of Republican 
    dissimilarity across a range of target distributions with different 
    compactness parameters $\rho$.}
  \label{fig:fl50-valid}
\end{figure}

For each target distribution, we sample 1,500 redistricting plans from both the
SMC and MCMC algorithms, and repeat the sampling four times in order to produce
$\hat R$ estimates. The MCMC algorithm is 
initialized with a random SMC-drawn map and first run for 500 warm-up iterations.
To validate and compare the samples, we reweight the enumerated redistricting
plans by $\tau(\xi)^\rho$, and then produce quantile-quantile plots of the
Republican dissimilarity index, which are shown in Figure~\ref{fig:fl50-valid}.

Across the range of $\rho$ values, the agreement between the SMC sample and 
target distributions is excellent, even in the lower tail. The $\hat R$ values 
for the SMC algorithm were also all less than 1.003. The MCMC algorithm fares
well in general but has noticeable bias for $\rho=0.9$ and $\rho=1.2$, and
has some additional misses for the other target distributions as well, which
manifest as small protuberances in the quantile-quantile plot. The $\hat R$ 
values for the MCMC algorithm were all less than 1.004, indicating that the
overall location and scale of the Republican dissimilarity were estimated well,
but $\hat R$ is not designed to capture these smaller-scale deviations from the
target distribution.

Here and in Section~\ref{sec:validation}, we validated and compared the SMC and
MCMC algorithms with several summary statistics. This reflects the applied
use case for redistricting analysis.  But it is also informative to study how
well the SMC algorithm can target the actual distribution of plans themselves.
Of course, we cannot expect the algorithm to perform well in this regard if 
there are fewer samples then there are plans, which is the case in almost 
all real-world problems. So we subset the enumerated plans to those with 
$\dev(\xi)\le 0.01$, of which there are just 38. We first generate 10,000 
samples from the SMC algorithm, targeting a distribution with $\rho=1$.
We measure the discrepancy between the sampled distribution of the 38 plans and 
the enumerated set weighted to $\tau(\xi)$ with the total variation distance, 
which is 0.0178 for this sample. Increasing the sample size to 50,000 decreases
the total variation distance to 0.0149.

\section{Algorithm Implementation Details}
\label{app:implementation}

\subsection{Estimating $K_i$}
A natural approach to estimating $K_i$ is to draw a moderate number of spanning 
trees $\mathcal{T}_i \subseteq \mathcal{T}(\widetilde{G}_i)$ and compute
$ok(T)$ for each $T\in\mathcal{T}_i$. The sample maximum, or the
sample maximum plus some small buffer amount, would then be an
estimate of the true maximum $\widehat{K}_i$ and an appropriate choice
of $k_i$.  In practice, we find little noticeable loss in algorithmic
accuracy even if $k_i<K_i$.  The following proposition theoretically
justifies this finding.
As above, $d_e$ represents the population deviation of the
district induced by removing edge $e$ from a spanning tree.
\begin{prop}\label{prop:qksimp} \spacingset{1}
    The probability $q(e=e^*\mid\F)$, i.e., the probability 
    that an edge $e$ is selected to be cut at iteration 
    $i$, given that the tree $T$ containing $e$  has been drawn,
    and that $e$ would induce a valid district, satisfies \[
        \max\qty{0, q(d_e\le d_{e_{k_i}}\mid\F)\qty(1+\frac{1}{k_i})-1}
        \le q(e=e^*\mid\F) \le \frac{1}{k_i},
    \]
    where $\F$ is the $\sigma$-field generated by $\{T,\pop(V_i)\in[P_i^-,P_i^+]\}$. 
\end{prop}
\begin{proof}
    We can write 
    \begin{equation*}
      q(e=e^*\mid\F) \ = \ q(e=e^*, d_e\le d_{e_{k_i}}\mid\F)
      \ = \ \frac{1}{k_i}q(d_e\le d_{e_{k_i}}\mid\F), 
    \end{equation*}
    This holds because the edge $e$ will not be cut unless $d_e\le d_{e_{k_i}}$, 
    i.e., if $e$ is among the top $k_i$ edges. We then have immediately
    that $q(e=e^*\mid\F)\le k_i^{-1}$.  Additionally, using the lower 
    Fr\'echet inequality, we find the lower bound
    \begin{align*}
        q(e=e^*\mid\F) \ &= \ q(e=e^*, d_e\le d_{e_{k_i}}\mid\F) \\ 
        &\ge \ \max\qty{0, q(e=e^*\mid\F)+q(d_e\le d_{e_{k_i}}\mid\F)-1} \\
        &= \ \max\qty{0, \frac{1}{k_i}q(d_e\le d_{e_{k_i}}\mid\F)
            +q(d_e\le d_{e_{k_i}}\mid\F)-1} \\
        &= \ \max\qty{0, q(d_e\le d_{e_{k_i}}\mid\F)\qty(1+\frac{1}{k_i})-1}.
        \qedhere
    \end{align*}
\end{proof}

If $k_i\ge K_i$, then $q(e=e^*\mid\F)$ is exactly $k_i^{-1}$, a fact which 
is used in the proof of Proposition~\ref{lem:splitprob}.  This result, which is
proved using a simple Fr\'echet bound, shows that as long as
$q(d_e\le d_{e_{k_i}}\mid\F)$ is close to 1, using $k_i^{-1}$ in
Proposition~\ref{lem:splitprob} is a good approximation to the true
sampling probability.

Having sampled $\mathcal{T}_i$, we can compute for each value of $k$
the sample proportion of trees where a randomly selected edge $e$
among the top $k$ of edges of the tree is also among the top $k$ for
the other trees---in effect estimating $q(d_e\le d_{e_{k_i}}\mid\F)$.
We may then choose $k_i$ to be the smallest $k$ for which this
proportion exceeds a pre-set threshold (e.g., 0.99). We have found
that this procedure, repeated at the beginning of each sampling stage,
efficiently selects $k_i$ without compromising the ability to sample
from the target distribution.

\subsection{Calculating $\psi([\xi])$}

To calculate $\psi([\xi])$, we first observe that sequentially valid labelings
of a particular unlabeled plan are in bijective  correspondence with certain 
increasing sequences of connected subgraphs of the quotient graph $G/\xi$. 
Specifically, as noted in the text, for every $1\le i\le n-1$, the subgraph 
$A_i\dfeq \{i+1,i+2,\dots,n\}$ of $G/\xi$ is connected if and only if $\xi$ is 
a sequentially valid labeling. Let $j=n-i$; then the sequence $A_j$ is increasing. 
Going from $A_j$ to $A_{j+1}$ for any $j$ involves adding a vertex in $G/\xi$ 
which is adjacent to $A_j$. 
This fact provides an easy scheme to generate the number of sequentially valid 
labelings: for each vertex $v$ in $G/\xi$, let $A_1=\{v\}$.
Then pick a neighbor of $A_1$ and add it to the set to form $A_2$.
Continue in this fashion until $A_{n-1}$ contains all but one vertex of $G/\xi$.
Undoing our bijection, this remaining vertex is labeled 1; the vertex added
between $A_{n-2}$ and $A_{n-1}$ is labeled 2, and so on.

\begin{figure}[!t]
  \centering \spacingset{1}
  \includegraphics[width=6.5in]{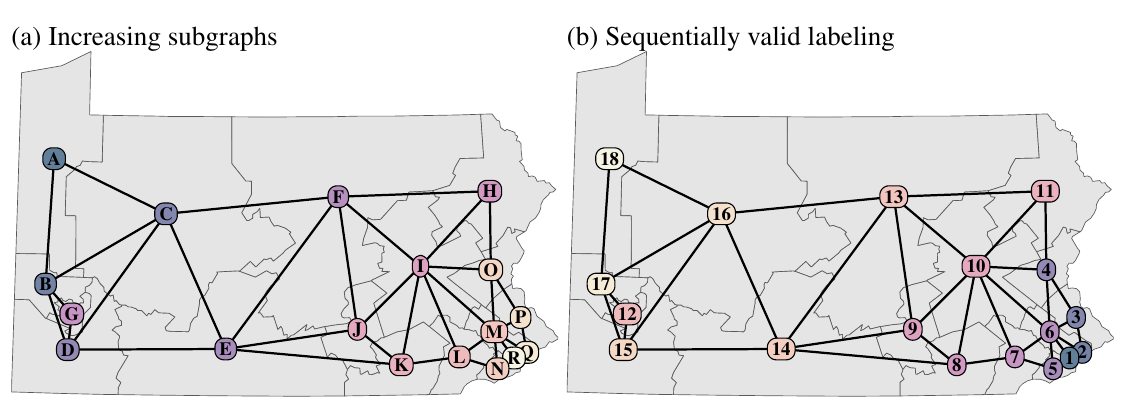}
  \caption{Schematic of the process to generate a sequentially valid labeling,
  using the district-level graph for the Pennsylvania court-imposed plan.
  Subgraphs are built in alphabetical order in panel (a): $\{A\}, \{A,B\}, \dots, \{A,B,\dots,Q\}$.
  This corresponds to a sequentially valid labeling shown in panel (b).}
  \label{fig:label-schem}
\end{figure}

Figure~\ref{fig:label-schem} illustrates this scheme on district-level graph 
for the Pennsylvania court-imposed plan.
We pick an arbitrary vertex, denoted by ``A" in Figure~\ref{fig:label-schem}(a). 
Then we pick a neighbor, denoted ``B," and add it to the subgraph.
We continue this way, adding the vertices indicated by the alphabetical ordering, 
until we have covered the whole district-level graph.
Then the last vertex added, ``R," is labeled 1 in the district labeling,
as shown in Figure~\ref{fig:label-schem}(b).
Vertex ``Q" is labeled 2, and so on, until vertex ``A" is labeled 18.
One can easily check that this labeling is sequentially valid---at every
point, the region of the map corresponding to unlabeled districts is contiguous.

As Section~\ref{subsubsec:psi} mentioned, we adopt different strategies for
calculating $\psi([\xi])$ when $n\le 13$ and $n>13$. 
When there are no more than 13 districts, we simply recurse down the tree of
possibilities for generating all sequentially valid plans.
In the example of Figure~\ref{fig:label-schem}, we have 18 options for the
first vertex in the subgraph. Once we have picked vertex ``A", then we could
add ``B" or ``C". If we add ``B", then we could add ``C", ``D", or ``G". 

Clearly this tree grows quite large very quickly.
Fortunately, there are many duplicate nodes---in our example,
adding vertex $B$ and then $C$ in sequence produces the same subgraph as if we
had added $C$ and then $B$.
Thus by memoizing our counting function we can significantly
reduce the number of tree branches we must explore in full.

When $n>13$, it is no longer computationally feasible to perform the recursion
for each sampled plan in what will often be a large number of sampled plans.
In these cases (it should be noted that for the 2020 redistricting cycle, only
nine states had more than 13 districts), we can only estimate the number of 
sequentially valid labelings.
To do so, we generate a large number of random sequentially valid labelings by
following the scheme outlined above, picking vertices of $G/\xi$ to add to each 
$A_j$ uniformly at random. Since we observe the number of candidate vertices to
add at each stage, and because each set of choices produces a unique sequentially
valid labeling, we can compute the probability of drawing each sequentially 
valid labeling directly. 

Denote the probability distribution created by this sampling scheme by 
$p_{\text{sv}}$, and the (not-probability) measure which assigns mass 1 to all
$n!$ relabelings by $\mu_{\text{all}}$. Our goal here is to estimate \[ 
    \int \ind\{\sigma\text{ is sequentially valid}\}\dd\mu_{\text{all}}(\sigma)
\] the total number of sequentially valid labelings. This can be done with our
sample from $p_{\text{sv}}$, since it is supported on precisely the same set
for which the indicator function takes a value of 1. We need only take the mean
of the inverse of the probability of each sampled labeling.

\begin{figure}[ht]
  \centering \spacingset{1}
  \includegraphics[width=6.5in]{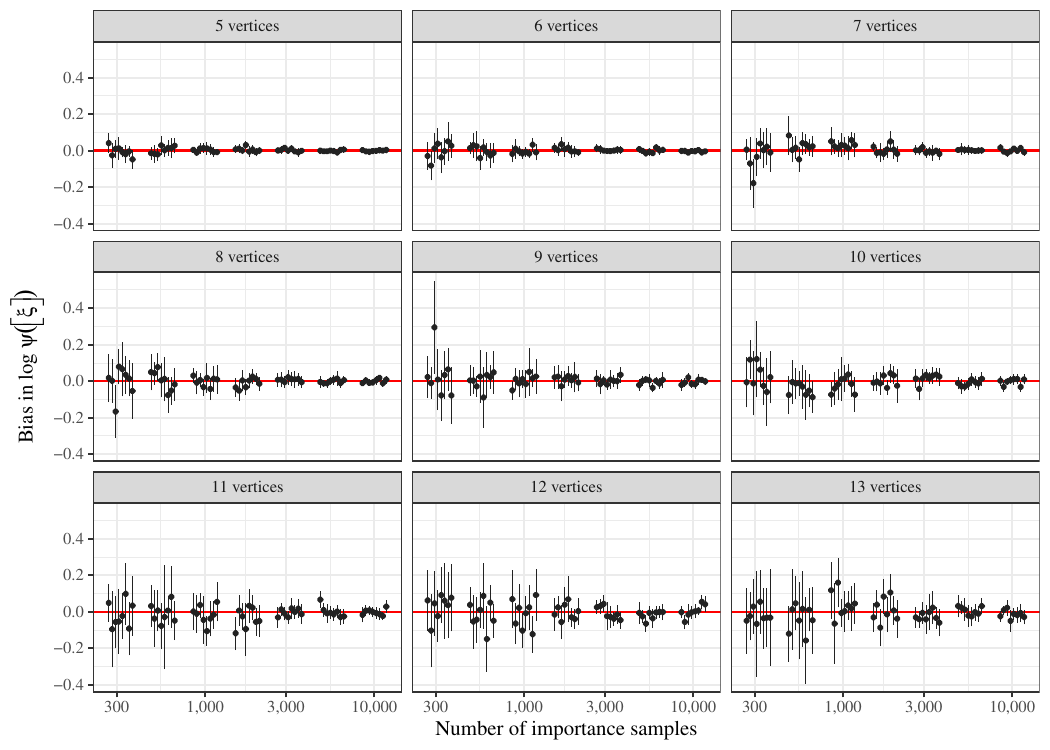}
  \caption{Bias in importance sampling estimates of $\log\psi([\xi])$ by the
  number of importance samples and the number of vertices in the district-level
  graph.  District-level graphs were selected as random subsets of the graph
  shown in \ref{fig:label-schem}. Eight replicate experiments were performed
  for each importance sample size and graph size. The vertical bars for each
  point span two standard errors in either direction. Standard errors were
  calculated with the delta method.}
  \label{fig:label-bias}
\end{figure}

This importance sampling estimate is quite accurate, since the proposal support 
matches the target support exactly, and there is not too much variation in the 
proposal probabilities. The estimate can be made arbitrarily good by increasing
the number of importance samples. 
This is demonstrated in Figure \ref{fig:label-bias}, which shows how the bias
in importance sampling estimates varies with the number of importance samples and
the overall size of the district-level graph.
In our software, we use on the order of 1,500
samples for $n=14$ by default, and more when $n$ is larger (e.g., around 3,000 
for the $n=28$ Florida congressional districts). These numbers are chosen to
ensure that the relative standard error (also known as the \textit{coefficient
of variation}) of the $\psi([\xi])$ estimate is controlled to a reasonable amount,
something we demonstrate next for the case of Pennsylvania.

For the district-level graph shown above in \ref{fig:label-schem}, we have
$\log\psi([\xi])=29.999$, calculated exactly using the recursive procedure 
described above. Using the importance sampling procedure with 2,000 samples, 
we estimate a value of $29.949$. This corresponds to a relative error of 0.051.
We can also calculate the relative standard error of this example estimate with the
delta method, which yields 0.052. In other words, the importance sampling error 
in the estimate of $\psi([\xi])$ is on the order of 5\%. 
This is also small on the scale of the variation in $\psi([\xi])$ across plans: 
for the six comparison plans used in the main text, 
$\log\psi([\xi])$ ranges from 28.456 to 30.317. 
So the error in the importance sampling estimate ($|29.949 - 29.999|=0.05$) 
is just 2.7\% of the range of $\log\psi([\xi])$ across these six plans.

Finally, we note that the procedures used here to calculate $\psi([\xi])$ can
also be applied to partial plans. Notating this formally is more difficult, but
a partial plan can be viewed as a (unbalanced) plan with fewer districts. 
Calculating $\psi$ for these partial plans and incorporating them into
the SMC weights in Algorithm~\ref{algo:smc} improves the sampling efficiency
without changing the target distribution, as long as each partial $\psi$'s 
contribution to the weights is canceled in the following iteration.
This is the approach taken by our software implementation.

\subsection{Stabilizing Importance Weights}

When $\rho\neq 1$ or when the constraints imposed by $J$ are severe,
there can be substantial variance in the importance sampling
weights. For large maps with $\rho=0$, for instance, the weights will generally 
span hundreds if not thousands of orders of magnitude. This reflects the
general computational difficulty in sampling uniformly from
constrained graph partitions. As \citet{najt2019} show, sampling of
node-balanced graph partitions is computationally intractable in the
worst case. In such cases, the importance sampling estimates will be
highly variable, and resampling based on these weights may lead to
degenerate samples with only one unique map.

When the importance weights are variable but not quite so extreme, we
find it useful to truncate the normalized final weights (such that their
mean is 1) from above at a value $w_{\max}$ at the end of sampling. The
theoretical basis for this maneuver is provided by
\citet{ionides2008}, who proved that as long as $w_{\max}\to\infty$
and $w_{\max}/S\to 0$ as $S\to\infty$, the resulting estimates are
consistent and have bounded variance (since the truncation occurs only
after the final SMC step, these conclusions, which were made in the
context of importance sampling, carry over.) One such choice we have found to
work well for the weights generated by this sampling process is
$w_{\max}=S^{0.4}/100$, though for particular maps other choices of
exponent and constant multiplier may be superior.

Truncation is no panacea, however.  As with any method that relies on
importance sampling, it is critical to examine the distribution of
importance weights to ensure that they will yield acceptable resamples.

\subsection{Computational Complexity}

The two asymptotically slowest steps of the SMC algorithm are
computing $\tau(G_i)$ for every district $G_i$ and drawing a spanning
tree using Wilson's algorithm for each iteration.  All other steps,
such as computing $d_e$ and
$|\C(G_i^{(j)},\widetilde{G}_i^{(j)})|$, are linear in the
number of vertices, and are repeated at most once per
iteration.\footnote{To compute $d_e$, we walk depth-first over the
  tree and store, for each node, the total population of that node and
  the nodes below it. This allows for $O(1)$ computation of $d_e$ for
  all edges.}  Computing $\tau(G_i)$ requires computing a determinant,
which currently has computational complexity $O(|V_i(\xi)|^{2.373})$
though most implementations are $O(|V_i(\xi)|^3)$.  Since this must be
done for each district of size roughly $m/n$, the total complexity for
sampling one plan is $O(n\cdot(m/n)^{2.373})$. For the spanning trees,
the expected runtime of Wilson's algorithm is the mean hitting time of
the graph, which is $O(m^2)$ in the worst case. So the total complexity 
for each sample is roughly $O(nm^2+m^{2.373}n^{-1.373})$ (ignoring the 
random rejection procedure).
Note that when $\rho=1$, we need not compute $\tau(G_i)$, and the
total complexity is roughly $O(nm^2)$.

\end{document}

%% file: header.tex
\usepackage{mathtools}
\usepackage{amsthm}
\usepackage{amssymb}
\usepackage{enumerate}
\usepackage{thmtools}
\usepackage{tikz}
\usepackage{xstring}
\usepackage{algorithm}
\usepackage{siunitx}
\usepackage{booktabs}
\usepackage[colorlinks,citecolor=darkgray,linkcolor=black]{hyperref}
\usepackage{caption}
\usepackage{subcaption}
\usepackage[italicdiff]{physics}
\usepackage[compact]{titlesec}
\usepackage{times}
\allowdisplaybreaks
\usepackage{natbib}
\newcommand\spacingset[1]{\renewcommand{\baselinestretch}%
  {#1}\small\normalsize}  
\theoremstyle{plain}
\declaretheorem[name=Proposition]{prop}

\newcommand\numberthis{\addtocounter{equation}{1}\tag{\theequation}}

\newcommand{\R}{\ensuremath{\mathbb{R}}}
\newcommand{\Z}{\ensuremath{\mathbb{Z}}}
\newcommand{\F}{\ensuremath{\mathcal{F}}}
\newcommand{\C}{\ensuremath{\mathcal{C}}}
\renewcommand{\empty}{\varnothing}

\newcommand{\dfeq}{\coloneqq}

\DeclareMathOperator{\E}{\mathbb{E}}

\newcommand{\ind}{\mathbf{1}}

\newcommand{\cvd}{\xrightarrow{\,d\,}}

\newcommand{\Norm}{\mathcal{N}\qty}

\DeclareMathOperator{\pop}{pop}
\DeclareMathOperator{\dev}{dev}
\DeclareMathOperator{\rem}{rem}
\DeclareMathOperator{\spl}{spl}

\newcommand{\appropto}{\mathrel{\vcenter{
  \offinterlineskip\halign{\hfil$##$\cr
    \propto\cr\noalign{\kern2pt}\sim\cr\noalign{\kern-2pt}}}}}